\documentclass{aa}

\usepackage{graphicx}

\usepackage{placeins} %
\usepackage{float}

\usepackage{txfonts}
\usepackage[dvipsnames]{xcolor}
\usepackage{gensymb} %
\usepackage{tikz}
\usetikzlibrary{decorations, decorations.text}
\usepackage{hyperref}
\hypersetup{
    colorlinks,
    citecolor=blue,
    linkcolor=blue
    }
\usepackage{subcaption}
\usepackage{pdfcomment}%

\begin{document} 

\title{Constraining the giant radio galaxy population with machine learning and Bayesian inference}
\authorrunning{Mostert \& Oei et al.}
\titlerunning{Giant radio galaxy constraints from machine learning and Bayesian inference}
   
\author{Rafa\"el I.J. Mostert\thanks{These authors contributed equally to this article.}\fnmsep\thanks{E-mail: mostert@strw.leidenuniv.nl, oei@strw.leidenuniv.nl}\fnmsep\inst{1,2} \and
Martijn S.S.L. Oei\footnotemark[1]\fnmsep\footnotemark[2]\fnmsep\inst{1,3} \and
B. Barkus\inst{4} \and
Lara Alegre\inst{5} \and
Martin J. Hardcastle\inst{6}\and
Kenneth J. Duncan\inst{5}\and
Huub J.A. R\"{o}ttgering\inst{1}\and
Reinout J. van Weeren\inst{1} \and
Maya Horton\inst{6}}

\institute{Leiden Observatory, Leiden University, Einsteinweg 55, 2333 CA Leiden, The Netherlands \and
ASTRON, the Netherlands Institute for Radio Astronomy, Oude Hoogeveensedijk 4, 7991 PD Dwingeloo, The Netherlands \and
Cahill Center for Astronomy and Astrophysics, California Institute of Technology, 1216 E California Blvd, CA 91125 Pasadena, USA \and
School of Physical Sciences, The Open University, Walton Hall, Milton Keynes, MK7 6AA, UK \and
Institute for Astronomy, Royal Observatory, Blackford Hill, Edinburgh, EH9 3HJ, UK \and
Centre for Astrophysics Research, Department of Physics, Astronomy and Mathematics, University of Hertfordshire, College Lane, Hatfield AL10 9AB, UK
}
\date{Received 10 December 2023; Accepted 23 February 2024}

  \abstract
  {
  Large-scale sky surveys at low frequencies, such as the LOFAR Two-metre Sky Survey (LoTSS), allow for the detection and characterisation of unprecedented numbers of giant radio galaxies (GRGs, or `giants', of at least $l_\mathrm{p,GRG} \coloneqq 0.7\ \mathrm{Mpc}$ long).
  This, in turn, enables us to study giants in a cosmological context.
  A tantalising prospect of such studies is a measurement of the contribution of giants to cosmic magnetogenesis.
  However, this measurement requires en masse radio--optical association for well-resolved radio galaxies and a statistical framework to infer GRG population properties.
    }
   {
   By automating the creation of radio--optical catalogues, we aim to significantly expand the census of known giants.
   With the resulting sample and a forward model that takes into account selection effects, we aim to constrain their intrinsic length distribution, number density, and lobe volume-filling fraction (VFF) in the Cosmic Web.
   }
   {
   We combined five existing codes into a single machine learning (ML)--driven pipeline that automates radio source component association and optical host identification for well-resolved radio sources.
   We created a radio--optical catalogue for the entire LoTSS Data Release 2 (DR2) footprint and subsequently selected all sources that qualify as possible giants.
   We combined the list of ML pipeline GRG candidates with an existing list of LoTSS DR2 crowd-sourced GRG candidates and visually confirmed or rejected all members of the merged sample.
   To infer intrinsic GRG properties from GRG observations, we developed further a population-based forward model and constrained its parameters using Bayesian inference.
   }
   {
   Roughly half of all GRG candidates that our ML pipeline identifies indeed turn out to be giants upon visual inspection, whereas the success rate is 1 in 11 for the previous best giant-finding ML technique in the literature.
   We confirm $5,647$ previously unknown giants from the crowd-sourced LoTSS DR2 catalogue and $2,597$ previously unknown giants from the ML pipeline.
   Our confirmations and discoveries bring the total number of known giants to at least $11,585$.
   Our intrinsic GRG population forward model provides a good fit to the data.
   The posterior indicates that the projected lengths of giants are consistent with a curved power law probability density function whose initial tail index $\xi(l_\mathrm{p,GRG}) = -2.8\pm0.2$ changes by $\Delta\xi = -2.4 \pm 0.3$ over the interval up to $l_\mathrm{p} = 5\ \mathrm{Mpc}$.
   We predict a comoving GRG number density $n_\mathrm{GRG} = 13 \pm 10\ (100\ \mathrm{Mpc})^{-3}$, close to a recent estimate of the number density of luminous non-giant radio galaxies.
   With the projected length distribution, number density, and additional assumptions, we derive a present-day GRG lobe VFF $\mathcal{V}_\mathrm{GRG-CW}(z=0) = 1.4 \pm 1.1 \cdot 10^{-5}$ in clusters and filaments of the Cosmic Web.
   }
   {
   We present a state-of-the-art ML-accelerated pipeline for finding giants, whose complex morphologies, arcminute extents, and radio-emitting surroundings pose challenges.
   Our data analysis suggests that giants are more common than previously thought.
   More work is needed to make GRG lobe VFF estimates reliable, but tentative results imply that it is possible that magnetic fields once contained in giants pervade a significant (${\gtrsim}10\%$) fraction of today's Cosmic Web.
   }


   \maketitle

\section{Introduction}
\label{sec:intro}
Recent radio Stokes $I$ imaging and rotation measure observations show that filaments of the Cosmic Web are magnetised \citep[e.g.][]{Govoni2019, deJong2022, Carretti2023} with $B \sim 10^0$--$10^2\ \mathrm{nG}$ \citep[e.g.][]{Vazza2021}.
However, the origin of these magnetic fields remains highly uncertain.
In a primordial magnetogenesis scenario \citep[e.g.][]{Subramanian2016}, the seeds of intergalactic magnetic fields can be traced to the Early Universe.
This scenario is not problem-free: primordial magnetic fields that arise before the end of inflation are typically too weak to match observations, while fields that arise after inflation (but before recombination) typically have coherence lengths that are too small.
Alternatively, in an astrophysical magnetogenesis scenario, the seeds of intergalactic magnetic fields are predominantly spread by energetic astrophysical phenomena in the more recent Universe, such as radio galaxies (RGs) and supernova-driven winds \citep[e.g.][]{Vazza2017}.
In this latter scenario, giant radio galaxies (GRGs, or `giants') may play a significant role in the magnetisation of the intergalactic medium (IGM), as their associated jets can carry magnetic fields of strength $B \sim 10^2\ \mathrm{nG}$ from host galaxies to cosmological, megaparsec-scale distances \citep[e.g.][]{Oei2022Alcyoneus}.

Efforts to measure the contribution of giants to astrophysical magnetogenesis in filaments of the Cosmic Web have only recently begun, with the advent of systematically processed, sensitive, low-frequency sky surveys such as the Low Frequency Array \citep[LOFAR;][]{vanHaarlem2013} Two-metre Sky Survey \citep[LoTSS;][]{Shimwell2017}.
By carrying out both a manual search for giants in LoTSS DR2 \citep{Shimwell2022} pipeline products and a rigorous statistical analysis, \citet{Oei2023Distribution} inferred a key statistic: the volume-filling fraction (VFF) of GRG lobes within clusters and filaments of the Local Universe, $\mathcal{V}_\mathrm{GRG-CW}(z = 0)$.
However, much uncertainty remains as to its value, which is set by the GRG number density, GRG length distribution, and GRG length--lobe volume relation.

As the number of observed radio galaxies rapidly increases with decreasing angular length, the time required to manually associate radio source components and identify host galaxies in the optical logically increases.
Machine learning (ML)--based methods have the potential to massively accelerate the detection of various radio source classes by complementing and eventually replacing manual methods \citep[e.g.][]{Proctor2016, Gheller2018, Lochner2021, Mostert2023remnants}.
The potential for detecting giants was demonstrated by \citet{Dabhade2020October}, who visually inspected the $1,600$ ML-identified GRG candidates of \citet{Proctor2016} and thereby discovered $151$ giants.
By combining into a single pipeline multiple ML-based and rule-based algorithms that automate both the radio component association and the optical host galaxy identification, we aim to improve upon the $9\%$ precision of \citet{Proctor2016}'s ML predictions.

As part of the present study, we constructed a LoTSS DR2 GRG sample of unparalleled size, by combining results from a visual search by astronomers \citep{Oei2023Distribution}, a visual search by citizen scientists \citep{Hardcastle2023}, and a ML-accelerated search (this article; Sect.~\ref{sec:methods}).
With a definitive LoTSS DR2 GRG sample in hand, we refined the Bayesian forward model presented in \citet{Oei2023Distribution}, and finally constrained several key geometric quantities pertaining to giants.

In Sect.~\ref{sec:theory}, we briefly review, generalise, and develop the statistical GRG geometry theory of \citet{Oei2023Distribution}.
In Sect.~\ref{sec:data}, we introduce the LoTSS DR2 data in which we searched for giants.
In Sect.~\ref{sec:methods}, we describe the methods that we used to build our definite LoTSS DR2 GRG sample, and explain how we used the theory of Sect.~\ref{sec:theory} in practice to infer GRG quantities of interest.
In Sect.~\ref{sec:results}, we present our findings regarding the projected proper length distribution for giants, their comoving number density, and their instantaneous lobe VFF in clusters and filaments of the Cosmic Web.
In Sect.~\ref{sec:discussion}, we discuss caveats of the present work, compare our results with previous results, and propose directions for future work, before we conclude in Sect.~\ref{sec:conclusions}.

We assume a flat $\mathrm{\Lambda CDM}$ model with
parameters from \citet{Planck2020}: $h=0.6766$, $\Omega_\mathrm{BM,0}=0.0490$, $\Omega_\mathrm{M,0}=0.3111$, and $\Omega_\mathrm{\Lambda,0}=0.6889$,
where $H_0 := h \cdot 100$ $\mathrm{km}$ $\mathrm{s}^{-1}$ $\mathrm{Mpc}^{-1}$.
We define giants as radio galaxies with a projected proper\footnote{In Cosmic Web filament environments, where giants appear most common \citep{Oei2024GiantsWeb}, lobes may expand along the Hubble flow, rendering their proper and comoving extents different. To avoid ambiguity, we stress that our projected lengths are proper instead of comoving. A less precise synonym for `projected proper length' often found in the literature is `largest linear size'.} length $l_\mathrm{p} \geq l_\mathrm{p,GRG} \coloneqq 0.7\ \mathrm{Mpc}$.
We define the spectral index $\alpha$ such that it relates to flux density $F_\nu$ at frequency $\nu$ as $F_\nu \propto \nu^\alpha$; under this convention, $\alpha < 0$ at radio frequencies for which synchrotron self-absorption is negligible. 

\section{Theory}
\label{sec:theory}
To infer the intrinsic length distribution, number density, and lobe VFF of giants, we used a Bayesian forward modelling approach that incorporates selection effects.
We adopt the framework described in \citet{Oei2023Distribution}, but generalise a few key formulae.
Furthermore, in a change that allows for the extraction of tighter parameter constraints from the data, we now predict joint projected proper length--redshift histograms rather than projected proper length distributions.

\subsection{RG total and projected proper lengths}
The central geometric quantity predicted by models of radio galaxy evolution \citep[e.g.][]{Turner2015, Hardcastle2018} is, simply, the RG's intrinsic proper length $l$.
Once the probability distribution of the intrinsic proper length random variable (RV) $L$ is known, one can estimate other geometric quantities of interest, such as the VFF of RG lobes in the Cosmic Web.
However, for the vast majority of observed RGs only a projected proper length $l_\mathrm{p}$ is available, as accurate measurements of jet inclination angles $\theta$ are currently challenging.
In order to fit statistical models to data from surveys such as LoTSS DR2, models should therefore predict the distribution of the projected proper length RV $L_\mathrm{p}$.

\subsection{GRG projected proper length: General}
We now show, first without adopting a specific parametric form for the distribution of $L$, how the cumulative density function (CDF) and probability density function (PDF) of the GRG projected proper length RV $L_\mathrm{p}\ \vert\ L_\mathrm{p} \geq l_\mathrm{p,GRG}$ can be calculated.
In particular, we suppose that $L$ has support from some length $l_\mathrm{min} \geq 0$ onwards.
Then $L_\mathrm{p} = L \sin{\Theta}$, where $\Theta$ is the inclination angle RV.
Assuming that, at least on cosmological scales, all RG orientations in three dimensions are equally likely,\footnote{This assumption is admissible, because even if the relative orientations of RGs and filaments are not random \citep[e.g.][]{Beckmann12024, Codis12018}, the uniformity of filament orientations on large scales leads to uniform RG orientations on large scales.} the CDF of $L_\mathrm{p}$ relates to the PDF of $L$ via
\begin{align}
    F_{L_\mathrm{p}}(l_\mathrm{p}) =
    \begin{cases}
        0 & \text{if } l_\mathrm{p} \leq 0;\\
        1 - \int_{l_\mathrm{min}}^\infty \sqrt{1 - \left(\frac{l_\mathrm{p}}{l}\right)^2}\ f_L(l)\ \mathrm{d}l & \text{if } 0 < l_\mathrm{p} \leq l_\mathrm{min};\\
        1 - \int_{l_\mathrm{p}}^\infty \sqrt{1 - \left(\frac{l_\mathrm{p}}{l}\right)^2}\ f_L(l)\ \mathrm{d}l & \text{if } l_\mathrm{p} > l_\mathrm{min}.\\
    \end{cases}
\label{eq:projectedLengthCDF}
\end{align}
We note that, in the usual scenario of $l_\mathrm{min} = 0$, the second case disappears.
Equation~\ref{eq:projectedLengthCDF} generalises Eq.~A.8 from \citet{Oei2023Distribution}; its derivation closely follows the one presented there.

The CDF of the GRG projected proper length RV $L_\mathrm{p}\ \vert\ L_\mathrm{p} \geq l_\mathrm{p,GRG}$ is
\begin{align}
    F_{L_\mathrm{p}\ \vert\ L_\mathrm{p} \geq l_\mathrm{p,GRG}}(l_\mathrm{p}) =
    \begin{cases}
    0 & \text{if } l_\mathrm{p} < l_\mathrm{p,GRG};\\
    1 - \frac{\int_{l_\mathrm{p}}^\infty \sqrt{1 - \left(\frac{l_\mathrm{p}}{l}\right)^2}\ f_L(l)\ \mathrm{d}l}{\int_{l_\mathrm{p,GRG}}^\infty \sqrt{1 - \left(\frac{l_\mathrm{p,GRG}}{l}\right)^2}\ f_L(l)\ \mathrm{d}l} & \text{if } l_\mathrm{p} \geq l_\mathrm{p,GRG}.
    \end{cases}
\end{align}
This result follows from combining Eq.~\ref{eq:projectedLengthCDF} and Eq.~A.12 from \citet{Oei2023Distribution}.\footnote{We have also assumed that $l_\mathrm{p,GRG} > l_\mathrm{min}$, which is the obvious case to consider.}
As PDFs follow from CDFs by differentiation, we find that the PDFs of $L_\mathrm{p}$ and $L_\mathrm{p}\ \vert\ L_\mathrm{p} \geq l_\mathrm{p,GRG}$ are related by
\begin{align}
    f_{L_\mathrm{p}\ \vert\ L_\mathrm{p} \geq l_\mathrm{p,GRG}}(l_\mathrm{p}) =
\begin{cases}
    0 & \text{if } l_\mathrm{p} < l_\mathrm{p,GRG};\\    
    \frac{f_{L_\mathrm{p}}(l_\mathrm{p})}{\int_{l_\mathrm{p,GRG}}^\infty \sqrt{1 - \left(\frac{l_\mathrm{p,GRG}}{l}\right)^2}\ f_L(l)\ \mathrm{d}l} & \text{if } l_\mathrm{p} \geq l_\mathrm{p,GRG}.
\end{cases}
\label{eq:projectedLengthPDFGRG}
\end{align}
We note that, throughout the support of $L_\mathrm{p}\ \vert\ L_\mathrm{p} \geq l_\mathrm{p,GRG}$, $f_{L_\mathrm{p}\ \vert\ L_\mathrm{p} \geq l_\mathrm{p,GRG}}(l_\mathrm{p})$ and $f_{L_\mathrm{p}}(l_\mathrm{p})$ are directly proportional -- the quantity in the denominator of Eq.~\ref{eq:projectedLengthPDFGRG} is merely a normalisation constant.\footnote{This is an example of a more general rule: for any RV $X$,
\begin{align}
    f_{X\ \vert\ X \geq y}(x) = \begin{cases}
        0 & \text{if }x < y;\\
        \frac{f_X(x)}{1 - F_X(y)} & \text{if } x \geq y.
    \end{cases}
\end{align}}

To find $f_{L_\mathrm{p}}(l_\mathrm{p})$ if $l_\mathrm{p} > l_\mathrm{min}$, it is helpful to perform a change of variables.
By defining $\eta \coloneqq \frac{l}{l_\mathrm{p}}$, we rewrite
\begin{align}
    F_{L_\mathrm{p}}(l_\mathrm{p}) = 1 - l_\mathrm{p}\int_1^\infty \sqrt{1 - \frac{1}{\eta^2}}\ f_L(l_\mathrm{p} \eta)\ \mathrm{d}\eta\ \ \ \ \text{if } l_\mathrm{p} > l_\mathrm{min}.
\end{align}
This form has the advantage that -- within the integral -- $l_\mathrm{p}$ occurs only in the integrand, whereas the form of Eq.~\ref{eq:projectedLengthCDF} features $l_\mathrm{p}$ in both the integrand and in the lower integration limit.
By differentation,
\begin{align}
    f_{L_\mathrm{p}}(l_\mathrm{p}) = &-\int_1^\infty \sqrt{1 - \frac{1}{\eta^2}}\ f_L(l_\mathrm{p} \eta)\ \mathrm{d}\eta\nonumber\\
    &-l_\mathrm{p}\int_1^\infty \sqrt{1 - \frac{1}{\eta^2}}\ \frac{\mathrm{d}f_L(l_\mathrm{p}\eta)}{\mathrm{d}l_\mathrm{p}}\ \mathrm{d}\eta\ \ \ \ \text{if } l_\mathrm{p} > l_\mathrm{min}.
\end{align}
To arrive at concrete expressions for the GRG projected proper length PDF of Eq.~\ref{eq:projectedLengthPDFGRG}, we must choose a specific parametric form for the distribution of $L$ or $L_\mathrm{p}$.

\subsection{GRG projected proper length: Curved power law}
\label{sec:theoryCurvedPowerLaw}
\citet{Oei2023Distribution} show that models that assume a Paretian tail for the RG intrinsic proper length distribution, and that include angular and surface brightness selection effects, can tightly reproduce the observed GRG projected proper length distribution.
The PDF of a Pareto-distributed RV is a simple power law, which is fully specified by a lower cut-off $l_\mathrm{min}$ and a tail index $\xi$.
However, there is a good reason to believe that the true GRG projected proper length PDF deviates from simple power law behaviour.
The true RG projected proper length PDF $f_{L_\mathrm{p}}$ will peak around a value set by the typical jet power, environment, lifetime, and inclination angle (amongst other properties).
Below this value, $f_{L_\mathrm{p}}$ will necessarily be an increasing function of $l_\mathrm{p}$; above this value, $f_{L_\mathrm{p}}$ will be a decreasing function.\footnote{This line of reasoning implicitly assumes that the distribution of $L_\mathrm{p}$ is unimodal.}
As giants embody the large-length tail of the distribution of $L_\mathrm{p}$, it is likely that the slope of $f_{L_\mathrm{p}\ \vert\ L_\mathrm{p} \geq l_\mathrm{p,GRG}}(l_\mathrm{p})$ first becomes more negative (and later becomes less negative) as $l_\mathrm{p}$ increases.

To remain close to the seemingly effective Pareto assumption of \citet{Oei2023Distribution}, we assume in this work that, at least for $l_\mathrm{p} \geq l_\mathrm{p,GRG}$, the RG projected proper length PDF is a curved power law:
\begin{align}
    f_{L_\mathrm{p}}(l_\mathrm{p}) \propto \left(\frac{l_\mathrm{p}}{l_\mathrm{p,GRG}}\right)^{\xi(l_\mathrm{p})}\ \ \ \ \text{if } l_\mathrm{p} \geq l_\mathrm{p,GRG},
\label{eq:RGProjectedProperLength}
\end{align}
where the exponent
\begin{align}
\xi(l_\mathrm{p}) \coloneqq \xi(l_\mathrm{p,1}) + \frac{l_\mathrm{p} - l_\mathrm{p,1}}{l_\mathrm{p,2} - l_\mathrm{p,1}}\left(\xi(l_\mathrm{p,2}) - \xi(l_\mathrm{p,1})\right)
\label{eq:exponentFunction}
\end{align}
is a linear function of $l_\mathrm{p}$.
As long as $l_\mathrm{p,1} \neq l_\mathrm{p,2}$, both projected proper length constants can be chosen arbitrarily; however, $l_\mathrm{p,1} \coloneqq l_\mathrm{p,GRG}$ seems to be a natural choice.
Adopting this choice, and defining $\Delta\xi \coloneqq \xi(l_\mathrm{p,2}) - \xi(l_\mathrm{p,1})$, leads to the final exponent formula
\begin{align}
    \xi(l_\mathrm{p}) = \xi(l_\mathrm{p,GRG}) + \frac{l_\mathrm{p} - l_\mathrm{p,GRG}}{l_\mathrm{p,2} - l_\mathrm{p,GRG}}\Delta\xi.
\label{eq:exponentFunction2}
\end{align}
We adopted $\xi(l_\mathrm{p,GRG})$ and $\Delta\xi$ as two parameters of our model.
We furthermore chose $l_\mathrm{p,2} \coloneqq 5\ \mathrm{Mpc}$, which is close to the largest currently known radio galaxy projected proper length \citep{Oei2022Alcyoneus, Oei2023Distribution}.
Being the first-order Taylor polynomial of an arbitrary function $\xi(l_\mathrm{p})$ at $l_\mathrm{p,GRG}$, Eq.~\ref{eq:exponentFunction} represents a natural generalisation of the constant tail index assumption of \citet{Oei2023Distribution}.
In particular, if model parameter $\Delta\xi = 0$, we recover the earlier Paretian model.

By the same reasoning as before, we find that if the RG projected proper length PDF is a curved power law for $l_\mathrm{p} \geq l_\mathrm{p,GRG}$, then the GRG projected proper length PDF is also a curved power law over this range:
\begin{align}
    f_{L_\mathrm{p}\ \vert\ L_\mathrm{p} \geq l_\mathrm{p,GRG}}(l_\mathrm{p}) \propto \left(\frac{l_\mathrm{p}}{l_\mathrm{p,GRG}}\right)^{\xi(l_\mathrm{p})}\ \ \ \ \text{if } l_\mathrm{p} \geq l_\mathrm{p,GRG}.
\label{eq:GRGProjectedProperLengthCurved}
\end{align}
The factors required to normalise $f_{L_\mathrm{p}}(l_\mathrm{p})$ and $f_{L_\mathrm{p}\ \vert\ L_\mathrm{p} \geq l_\mathrm{p,GRG}}(l_\mathrm{p})$ can be obtained numerically.

Whereas \citet{Oei2023Distribution} parametrised $f_L(l)$ and derived $f_{L_\mathrm{p}}(l_\mathrm{p})$ and $f_{L_\mathrm{p}\ \vert\ L_\mathrm{p} \geq l_\mathrm{p,GRG}}(l_\mathrm{p})$, we now parametrise $f_{L_\mathrm{p}}(l_\mathrm{p})$ and derive only $f_{L_\mathrm{p}\ \vert\ L_\mathrm{p} \geq l_\mathrm{p,GRG}}(l_\mathrm{p})$.
It is possible to start modelling at the level of $f_L(l)$, also in the context of curved power law PDFs, but the resulting expressions for $f_{L_\mathrm{p}}(l_\mathrm{p})$ and $f_{L_\mathrm{p}\ \vert\ L_\mathrm{p} \geq l_\mathrm{p,GRG}}(l_\mathrm{p})$ become tedious and rather uninsightful.
For simplicity, we therefore choose to parametrise $f_{L_\mathrm{p}}(l_\mathrm{p})$; we explore the alternative set-up in Appendix~\ref{app:theory}.

\subsection{GRG observed projected proper length}
\label{sec:theorySelection}
Equation~\ref{eq:GRGProjectedProperLengthCurved} describes a distribution of GRG projected proper lengths in the absence of observational selection effects.
Unfortunately, this distribution cannot be directly tested against GRG samples obtained from surveys, which are always affected by selection.
For a thorough description and derivation of selection effect modelling in the context of our framework, we refer the reader to Sect.~2.8 and Appendix~A.8 of \citet{Oei2023Distribution}; here, we shall only briefly introduce the expressions that we require.

A key result, adopted from Eq.~21 of \citet{Oei2023Distribution}, is that the GRG observed projected proper length RV $L_\mathrm{p,obs}\ \vert\ L_\mathrm{p,obs} \geq l_\mathrm{p,GRG}$ can be expressed as
\begin{align}
&f_{L_\mathrm{p,obs}\ \vert\ L_\mathrm{p,obs} \geq l_\mathrm{p,GRG}}(l_\mathrm{p}) =\begin{cases}
0 & \text{if } l_\mathrm{p} < l_\mathrm{p, GRG};\\
\frac{C\left(l_\mathrm{p},z_\mathrm{max}\right)f_{L_\mathrm{p}}\left(l_\mathrm{p}\right)}{\int_{l_\mathrm{p,GRG}}^\infty C\left(l_\mathrm{p}',z_\mathrm{max}\right)f_{L_\mathrm{p}}\left(l_\mathrm{p}'\right)\ \mathrm{d}l_\mathrm{p}'} & \text{if } l_\mathrm{p} \geq l_\mathrm{p, GRG},
\end{cases}
\label{eq:GRGObservedProjectedProperLength}
\end{align}
where $C$ is the completeness function.
More precisely, $C(l_\mathrm{p}, z_\mathrm{max})$ denotes the fraction of all RGs with projected proper length $l_\mathrm{p}$ in the volume up to cosmological redshift $z_\mathrm{max}$ that is detected and identified through the survey considered -- in this work, this will be LoTSS DR2.
The repeated factors in numerator and denominator reveal that, in order to compute $f_{L_\mathrm{p,obs}\ \vert\ L_\mathrm{p,obs} \geq l_\mathrm{p,GRG}}(l_\mathrm{p})$, we need to know $f_{L_\mathrm{p}}(l_\mathrm{p})$ up to a constant only (and on $l_\mathrm{p} \geq l_\mathrm{p,GRG}$ only).
More concerningly, we also see that selection effects that reduce the completeness by the same factor for all $l_\mathrm{p} \geq l_\mathrm{p,GRG}$ leave no imprint on $f_{L_\mathrm{p,obs}\ \vert\ L_\mathrm{p,obs} \geq l_\mathrm{p,GRG}}(l_\mathrm{p})$.
Therefore, such selection effects cannot be constrained by a GRG observed projected proper length analysis alone.

Under the assumption that the RG projected proper length PDF $f_{L_\mathrm{p}}(l_\mathrm{p})$ does not evolve between redshifts $z = z_\mathrm{max}$ and $z = 0$, the completeness function becomes
\begin{align}
C\left(l_\mathrm{p}, z_\mathrm{max}\right) = \frac{\int_0^{z_\mathrm{max}}p_\mathrm{obs}\left(l_\mathrm{p},z\right) r^2\left(z\right) E^{-1}\left(z\right)\ \mathrm{d}z}{\int_0^{z_\mathrm{max}}r^2\left(z\right)E^{-1}\left(z\right)\ \mathrm{d}z},
\label{eq:completeness}
\end{align}
where the observing probability $p_\mathrm{obs}(l_\mathrm{p},z)$ is the probability that an RG of projected proper length $l_\mathrm{p}$ at redshift $z$ is detected by a survey and its subsequent analysis steps (such as the machine learning pipeline considered in this work), $r$ denotes comoving radial distance, and $E(z)$ is the dimensionless Hubble parameter\footnote{In a \emph{flat} Friedmann--Lema\^itre--Robertson--Walker universe, the dimensionless Hubble parameter $E$ is
\begin{align}
    E\left(z\right) \coloneqq& \frac{H\left(z\right)}{H_0} = \sqrt{\Omega_\mathrm{R,0} \left(1+z\right)^4 + \Omega_\mathrm{M,0} \left(1+z\right)^3 + \Omega_\mathrm{\Lambda,0}}.
\end{align}}.
The appropriate form of $p_\mathrm{obs}(l_\mathrm{p},z)$ is determined by the selection effects relevant to the survey of interest and its analysis.

In this work, we consider GRG lobe surface brightness (SB) selection, which at present renders many members of the GRG population undetectable, and selection by limitations of our analysis, which causes in principle detectable giants to evade identification.
We described the former effect parametrically, and determined the latter effect empirically.
The effects yield functions $p_\mathrm{obs,SB}(l_\mathrm{p},z)$ and $p_\mathrm{obs,ID}(l_\mathrm{p},z)$, respectively, which then combine to form a single observing probability function through
\begin{align}
    p_\mathrm{obs}(l_\mathrm{p},z) = p_\mathrm{obs,SB}(l_\mathrm{p},z) \cdot p_\mathrm{obs,ID}(l_\mathrm{p},z).
\end{align}

\subsubsection{Selection effects: Surface brightness limit}
\label{sec:theorySurfaceBrightness}
RG lobes whose SBs are lower than some threshold $b_{\nu,\mathrm{th}}$, which typically equals the survey noise level $\sigma$ times a factor of order unity, cannot be detected.
Following Sect.~2.8.3 of \citet{Oei2023Distribution}, we modelled SB selection by assuming that the lobe SBs $B_\nu(\nu,l,z)$ at $\nu = \nu_\mathrm{obs}$ of RGs of intrinsic proper length $l = l_\mathrm{ref}$ residing at redshift $z = 0$ are lognormally distributed.
We parametrised $B_\nu(\nu_\mathrm{obs}, l_\mathrm{ref}, 0) = b_{\nu,\mathrm{ref}}\ S$, where $b_{\nu,\mathrm{ref}}$ is the median lobe SB and $S$ is a lognormally distributed RV with median $1$ and dispersion parameter $\sigma_\mathrm{ref}$.
The observing probability due to SB selection then is
\begin{align}
p_\mathrm{obs,SB}\left(l_\mathrm{p},z\right) &= \int_{s_\mathrm{min}}^\infty\sqrt{1 - \left(\frac{s_\mathrm{min}}{s}\right)^{-\frac{2}{\zeta}}}f_S\left(s\right) \mathrm{d}s;\label{eq:probabilityObservedSB}\\
    s_\mathrm{min} &= \frac{b_{\nu,\mathrm{th}}}{b_{\nu,\mathrm{ref}}}\left(\frac{l_\mathrm{p}}{l_\mathrm{ref}}\right)^{-\zeta}\left(1+z\right)^{3-\alpha};\\
    f_S\left(s\right) &= \frac{1}{\sqrt{2\pi}\sigma_\mathrm{ref} s}\exp{\left(-\frac{\ln^2s}{2\sigma_\mathrm{ref}^2}\right)}.
\end{align}
Here, $\alpha$ is the typical RG lobe spectral index, which we assumed fixed at $\alpha = -1$.
The exponent $\zeta$ determines how the SB distribution scales with projected proper length $l_\mathrm{p}$.

In a departure from \citet{Oei2023Distribution}, we did not fix $\zeta = -2$, but rather left $\zeta$ a free parameter which we fitted to the data.
Deviations from $\zeta = -2$ occur in at least two cases: when giant growth is not shape-preserving, and if the radio luminosity distributions of giants of different $l_\mathrm{p}$ are distinct.
Dynamical models of RGs in general predict that both cocoons \citep[e.g. Fig. 4 of][]{Turner2015} and lobes \citep[e.g. Fig. 9 of][]{Hardcastle2018} change shape over time, and in a jet power--dependent way.
There remains considerable uncertainty as to how shapes change throughout the giant phase: axial ratio--like measures generally show that RG lobes become more elongated during growth, but this trend could possibly reverse for giants, whose lobes might protrude from the clusters and filaments in which they are born.
Simulations suggest that, for such protrusions, the usual constant power-law profile assumptions for the ambient baryon density and temperature break down \citep[e.g. Fig. 8 of][]{Gheller2019}.
If lobes of giants widen over time, then $\zeta$ would decrease.
The second case occurs if the end-of-life lengths of RGs increase with jet power, so that the subpopulation that survives up to some $l_\mathrm{p}$ has its jet power distribution -- and therefore its radio luminosity distribution -- shifted upwards with respect to subpopulations at smaller $l_\mathrm{p}$.
This effect, which appears plausible given models \citep[e.g. Fig. 8 of][]{Hardcastle2018}, would increase $\zeta$.
At present, it seems hard to predict the net result on $\zeta$ of these counteracting effects.

\subsubsection{Selection effects: Non-identification}
\label{sec:theoryIdentification}
Every present-day survey search method (such as visual inspection by scientists, visual inspection by citizen scientists, and ML-based approaches) will fail to identify some giants that are in principle identifiable (in the sense that they lie above the detection threshold set by the noise).
For automated approaches, such as the ML-based approach presented in this work, identification can become more challenging for larger angular lengths $\phi$: one reason being the increased number of unrelated, interloping radio sources that cover the solid angle occupied by the RG.
We call the probability that an identifiable RG is indeed identified -- and therefore becomes part of the final sample -- $p_\mathrm{obs,ID}$.

\begin{figure}
    \centering
    \def\scaleFactor{.75}
    \begin{tikzpicture}[fill opacity=0.5, draw opacity = 0.25]
    \fill[CadetBlue!7!white] (0,0) circle(3.6*\scaleFactor);
    \fill[CadetBlue!14!white] (0,-.3*\scaleFactor) circle(3.2*\scaleFactor);

    \fill[CadetBlue!50!white](0,.5*\scaleFactor) circle(1.85*\scaleFactor);
    \fill[CadetBlue!50!white] (-0.9899*\scaleFactor,-1.49899*\scaleFactor) circle(1.45*\scaleFactor);
    \fill[CadetBlue!50!white] (0.9899*\scaleFactor,-1.49899*\scaleFactor) circle(1.45*\scaleFactor);
    
    \fill[CadetBlue!21!white](0,.5*\scaleFactor) circle(1.8*\scaleFactor);

\draw[color=CadetBlue!7!white, rotate=-90, postaction={decorate, decoration={text along path, raise=-8pt, text align={align=center}, text color=CadetBlue!50!white, text={|\tiny|All}, reverse path}}] (0,0) circle (3.6*\scaleFactor);
\draw[color=CadetBlue!14!white, rotate=-90, postaction={decorate, decoration={text along path, raise=-8pt, text align={align=center}, text color=CadetBlue!75!white, text={|\tiny|Identifiable}, reverse path}}] (0.3*\scaleFactor,0) circle (3.2*\scaleFactor);
\draw[color=CadetBlue!50!white, line width=0pt, rotate=-90, postaction={decorate, decoration={text along path, raise=-8pt, text align={align=center}, text color=CadetBlue!100!white, text={|\tiny|Identified}, reverse path}}] (-0.5*\scaleFactor,0) circle (1.85*\scaleFactor);
    
    \fill[CadetBlue!30!white] (-0.9899*\scaleFactor,-1.49899*\scaleFactor) circle(1.4*\scaleFactor);
    \fill[CadetBlue!30!white] (0.9899*\scaleFactor,-1.49899*\scaleFactor) circle(1.4*\scaleFactor);
    \draw[dashed, CadetBlue!50!white, thin] (0,.5*\scaleFactor) circle(1.8*\scaleFactor);
    \draw[dashed, CadetBlue!50!white, thin] (-0.9899*\scaleFactor,-1.49899*\scaleFactor) circle(1.4*\scaleFactor);
    \draw[dashed, CadetBlue!50!white, thin] (0.9899*\scaleFactor,-1.49899*\scaleFactor) circle(1.4*\scaleFactor);
    
    \node at (0,-3.2*\scaleFactor) (G) {\small \textcolor{CadetBlue}{$\mathcal{G} = \mathcal{G}_1 \cup \mathcal{G}_2 \cup \mathcal{G}_3$}};
    \node at (0,.5*\scaleFactor) (D) {\small \textcolor{CadetBlue}{$\mathcal{G}_1$}};
    \node at (-0.9899*\scaleFactor,-1.49899*\scaleFactor) (E) {\small \textcolor{CadetBlue}{$\mathcal{G}_2$}};
    \node at (0.9899*\scaleFactor,-1.49899*\scaleFactor) (F) {\small \textcolor{CadetBlue}{$\mathcal{G}_3$}};

\end{tikzpicture}
    \caption{
    Schematic of a three-method search for giants.
    Of all giants in the survey footprint up to $z = z_\mathrm{max}$, only those for which the lobe surface brightness at the observing frequency $\nu_\mathrm{obs}$ is above detection threshold $b_{\nu,\mathrm{th}}$ are identifiable.
    $\mathcal{G}$ denotes the actually identified set of giants.
    $\mathcal{G}_1$, $\mathcal{G}_2$, and $\mathcal{G}_3$ are the subsets identified by each method individually.
    As an example, we shade $\mathcal{G}_2 \cup \mathcal{G}_3$, which has overlap with $\mathcal{G}_1$, and which can be used to measure $p_\mathrm{obs,ID,1}(l_\mathrm{p},z)$.
    }
    \label{fig:giantsVenn}
\end{figure}
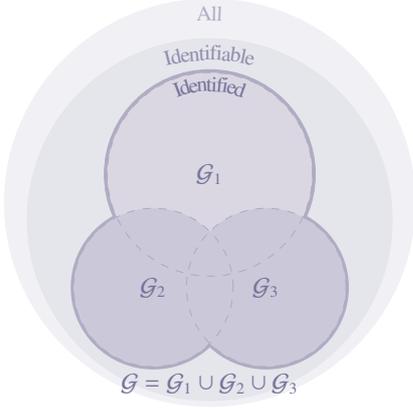

Say we have $M$ methods to search for giants in the same survey.
Let $\mathcal{G} = \{g_1, g_2, ..., g_N\}$ be the set of all identified giants (so that $|\mathcal{G}| = N$), and let $\mathcal{G}_i \subseteq \mathcal{G}$ be the subset identified by method $i$.
Figure~\ref{fig:giantsVenn} illustrates the set-up.
The projected proper length and cosmological redshift of giant $g$ are $l_\mathrm{p}(g)$ and $z(g)$, respectively.
To determine the identification probability $p_{\mathrm{obs,ID},i}(l_\mathrm{p},z)$ for method $i$, we first assume it to be of logistic form
\begin{align}
    p_{\mathrm{obs,ID},i}(l_\mathrm{p},z) = \frac{1}{1 + \exp{(-(\beta_{0,i} + \beta_{l_\mathrm{p},i} \cdot l_\mathrm{p} + \beta_{z,i} \cdot z))}}.
\end{align}
We obtain best-fit parameters $\hat{\beta}_{0,i}$, $\hat{\beta}_{l_\mathrm{p},i}$, and $\hat{\beta}_{z,i}$ by performing binary logistic regression with two explanatory variables on the data set $\mathcal{D}_i$, where
\begin{align}
    \mathcal{D}_i \coloneqq \left\{\left(\left[l_\mathrm{p}(g), z(g)\right], \mathbb{I}(g \in \mathcal{G}_i)\right)\ \vert\ g \in \bigcup_{j =1, j \neq i}^M \mathcal{G}_j\right\}.
\end{align}
The first element of each pair in $\mathcal{D}_i$ is a point in projected length--redshift space, whilst the second element is 0 or 1: $\mathbb{I}$ denotes the indicator function.
Qualitatively, $\mathcal{D}_i$ stores for each giant in the union of all GRG subsets except $\mathcal{G}_i$ its projected length--redshift coordinates, together with the success or failure of its identification through method $i$.

The implicit assumption here is that all $ g \in \bigcup_{j =1, j \neq i}^M \mathcal{G}_j$ are typical examples of identifiable giants at the relevant projected proper length and redshift.
We caution that this might not be true: giants with a peculiar morphology, or those lying in parts of the sky where optical identification is hard (e.g. towards the Galactic Plane or crowded regions of large-scale structure), may be identifiable in a radio surface brightness sense, but will nonetheless evade sample inclusion more often than other giants.
As a result, giants that \emph{do} end up in a sample -- such as $\bigcup_{j =1, j \neq i}^M \mathcal{G}_j$ -- will have more regular morphologies than giants in general and will lie in regions of the sky where optical identification is easier than for giants in general.
Typically, such giants are also more likely to be found by method $i$, and as a result our approach will probably render $p_{\mathrm{obs,ID},i}$ biased high.

Given a set of $M$ functions $\{p_{\mathrm{obs,ID},i}(l_\mathrm{p},z)\ \vert\ i \in \{1, 2, ..., M\}\}$, several possibilities exist to combine them into a single $p_\mathrm{obs,ID}(l_\mathrm{p},z)$.
At the minimum, $p_\mathrm{obs,ID}(l_\mathrm{p},z)$ is given by a point-wise maximum:
\begin{align}
    p_\mathrm{obs,ID}(l_\mathrm{p},z) = \max_{i \in \{1, 2, ..., M\}} p_{\mathrm{obs,ID},i}(l_\mathrm{p},z),
\label{eq:identificationProbabilityCombinationMaximum}
\end{align}
which is appropriate if methods tend to find the same identifiable giants -- as in our case.\footnote{
In case methods tend to find independent subsets of identifiable giants,
\begin{align}
    p_\mathrm{obs,ID}(l_\mathrm{p},z) = 1 - \prod_{i=1}^M (1 - p_{\mathrm{obs,ID},i}(l_\mathrm{p},z)).
\end{align}
We note that it is possible to design methods that find subsets of identifiable giants that have even less overlap than independent subsets have.}

\subsection{GRG number density}
\label{sec:theoryNumberDensity}
The preceding theory allows us to find the intrinsic, comoving number density of giants, $n_\mathrm{GRG}$, if we know the observed number of giants within a solid angle of extent $\Omega$ and in the volume up to $z_\mathrm{max}$, $N_\mathrm{GRG,obs}(\Omega, z_\mathrm{max})$.
We assume that, up to this redshift, $n_\mathrm{GRG}$ remains constant.
We note that we cannot calculate $n_\mathrm{GRG}$ using Eq. 30 from \citet{Oei2023Distribution}: this equation assumes $\xi(l_\mathrm{p,1}) = \xi(l_\mathrm{p,2})$.
We derive a more general expression by first noting that the number of giants observed within a solid angle of extent $\Omega$ in the volume up to $z_\mathrm{max}$ and with projected proper lengths between $l_\mathrm{p}$ and $l_\mathrm{p} + \mathrm{d}l_\mathrm{p}$ is
\begin{align}
    \mathrm{d}N_\mathrm{GRG,obs}(l_\mathrm{p},\Omega,z_\mathrm{max}) &= \frac{\Omega}{4\pi}\ n_\mathrm{GRG}\ f_{L_\mathrm{p}\ \vert\ L_\mathrm{p} \geq l_\mathrm{p,GRG}}(l_\mathrm{p})\ \mathrm{d}l_\mathrm{p}\ \cdot\nonumber\\
    &\int_0^{z_\mathrm{max}} p_\mathrm{obs}(l_\mathrm{p},z)\ 4\pi r^2(z)\ \frac{\mathrm{d}r}{\mathrm{d}z}\ \mathrm{d}z.
    \label{eq:numberOfGiantsDifferential}
\end{align}
Because
\begin{align}
    N_\mathrm{GRG,obs}(\Omega, z_\mathrm{max}) = \int_{l_\mathrm{p,GRG}}^\infty \mathrm{d}N_\mathrm{GRG,obs}(l_\mathrm{p},\Omega,z_\mathrm{max}),
\end{align}
we find, by isolating $n_\mathrm{GRG}$, that
\begin{align}
\label{eq:GRGNumberDensity}
&n_\mathrm{GRG}(l_\mathrm{p,GRG}, z_\mathrm{max}) = \frac{H_0}{c}\ \frac{4\pi}{\Omega}\ N_\mathrm{GRG,obs}(\Omega, z_\mathrm{max})\ \cdot\\
&\left(\int_{l_\mathrm{p,GRG}}^\infty f_{L_\mathrm{p}\ \vert\ L_\mathrm{p} \geq l_\mathrm{p,GRG}}(l_\mathrm{p}) \int_0^{z_\mathrm{max}} p_\mathrm{obs}(l_\mathrm{p},z)\ 4\pi r^2(z)\ E^{-1}(z)\ \mathrm{d}z\ \mathrm{d}l_\mathrm{p}\right)^{-1}.\nonumber
\end{align}
This expression is valid also beyond the context of power law or curved power law PDFs $f_{L_\mathrm{p}\ \vert\ L_\mathrm{p} \geq l_\mathrm{p,GRG}}(l_\mathrm{p})$.
We remark that $n_\mathrm{GRG}$ can depend sensitively on $l_\mathrm{p,GRG}$, the projected proper length used to define giants.
In contrast to the approach of \citet{Oei2023Distribution}, in this work we do not calculate $n_\mathrm{GRG}$ in a step \emph{following} inference of the framework's parameters, but rather include it as a parameter to be constrained \emph{during} inference.

\subsection{GRG lobe volume-filling fraction}
To constrain the contribution of giants to astrophysical magnetogenesis, we wish to know the volume-filling fraction of their lobes in clusters and filaments of the Cosmic Web, $\mathcal{V}_\mathrm{GRG-CW}$.
This quantity may have changed over cosmic time; in this work, we calculate it at the present day.
First, we make the approximation that all GRG lobes (in the Local Universe) lie in clusters and filaments.
In addition, we model the general RG relation between the two-lobe proper volume RV $V$ and the projected proper length RV $L_\mathrm{p}$ as a power law with scatter:
\begin{align}
    V = V_\mathrm{GRG} \cdot \left(\frac{L_\mathrm{p}}{l_{\mathrm{p,GRG}}}\right)^\gamma \cdot X,
\label{eq:VFFModel}
\end{align}
where $V_\mathrm{GRG}$ is the mean two-lobe proper volume of an RG with a projected proper length $l_\mathrm{p,GRG}$, and $\gamma$ is a constant exponent.
Furthermore, $X$, which we take to be independent of $L_\mathrm{p}$, is a non-negative, dimensionless RV with a mean of unity and an otherwise arbitrary distribution.
In Sect.~\ref{sec:resultsVFF}, we present observations that indicate that this model is reasonable.
Under this model,
\begin{align}
    \mathbb{E}[V\ \vert\ L_\mathrm{p} \geq l_\mathrm{p,GRG}] = \frac{V_\mathrm{GRG}}{l_\mathrm{p,GRG}^\gamma} \cdot \mathbb{E}[L_\mathrm{p}^{\gamma}\ \vert\ L_\mathrm{p} \geq l_\mathrm{p,GRG}].
\end{align}
Finally, if GRG lobes are sufficiently small or giants sufficiently rare (or both), the probability that there exist overlapping GRG lobes will be low.
If, indeed, GRG lobes do not overlap, $\mathcal{V}_\mathrm{GRG-CW} \propto \mathbb{E}[V\ \vert\ L_\mathrm{p} \geq l_\mathrm{p,GRG}]$ and $\mathcal{V}_\mathrm{GRG-CW} \propto n_\mathrm{GRG}$, so that
\begin{align}
\mathcal{V}_\mathrm{GRG-CW}(z) = &\frac{\mathbb{E}[V\ \vert\ L_\mathrm{p} \geq l_\mathrm{p,GRG}](z)\ n_\mathrm{GRG}(z)\ (1+z)^3}{\mathcal{V}_\mathrm{CW}(z)},
\label{eq:VFF}
\end{align}
where $\mathcal{V}_\mathrm{CW}$ denotes the VFF of clusters and filaments combined.

\subsection{GRG angular lengths} \label{sec:theoryAnglengths}
An object's angular length $\phi$, projected proper length $l_\mathrm{p}$, and cosmological redshift $z$ are related through
\begin{align}
    \phi(l_\mathrm{p},z) = \frac{l_\mathrm{p}(1+z)}{r(z)}.
\end{align}
Due to the expansion of the Universe, there exists a minimum angular length for objects of a given projected proper length.
If one defines giants as RGs with projected proper lengths $l_\mathrm{p} \geq l_\mathrm{p,GRG} \coloneqq 0.7\ \mathrm{Mpc}$, as in this work, then all giants have an angular length $\phi \geq 1.3'$ \citep{Oei2023Distribution}.
This fact has important consequences for GRG search campaigns.
At the LoTSS resolution of $\theta_\mathrm{FWHM} = 6''$, it implies that giants are always resolved and span at least 13 resolution elements.
Therefore, to model the detectability of giants at this resolution, one must consider their surface brightness (profiles), rather than their flux densities.

\subsection{Inference}
\label{sec:theoryInference}
Finally, we describe how the framework's six free parameters $\vec{\theta} \coloneqq [\xi(l_\mathrm{p,GRG}), \Delta\xi, b_{\nu,\mathrm{ref}}, \sigma_\mathrm{ref},\zeta, n_\mathrm{GRG}]$ can be inferred from a data set containing a projected length and redshift for each observed giant.
In particular, we consider a rectangle in projected proper length--cosmological redshift parameter space, within which our model assumptions are expected to hold.
We partition this rectangle into $N_\mathrm{b}$ equiareal bins of width $\Delta l_\mathrm{p}$ and height $\Delta z$.
We denote the coordinates of bin $i$'s centre as $(l_{\mathrm{p},i}, z_i)$.

We binned the data to obtain a two-dimensional histogram.
The number of giants found in bin $i$, $N_i$, is an RV with a Poisson distribution: $N_i \sim \mathrm{Poisson}(\lambda_i)$.
Its expectation $\lambda_i$ depends on the model parameters $\vec{\theta}$.
Assuming that the $\{N_i\}$ are independent, the log-likelihood becomes
\begin{align}
    \ln{\mathcal{L}(\{N_i\}\ \vert\ \vec{\theta})} = \sum_{i=1}^{N_\mathrm{b}} N_i \ln{\lambda_i(\vec{\theta})} - \lambda_i(\vec{\theta}) - \ln{(N_i!)}.
    \label{eq:logLikelihood}
\end{align}
The last term on the right-hand side of Eq.~\ref{eq:logLikelihood} is the same for all $\vec{\theta}$, and need not be calculated if one is interested in $\mathcal{L}$ up to a global constant only.\footnote{If one includes this term, it only needs to be calculated once. For numerical stability, it is helpful to note that
\begin{align}
    -\sum_{i=1}^{N_\mathrm{b}} \ln{(N_i!)} = -\sum_{i=1}^{N_\mathrm{b}}\sum_{j=2}^{N_i}\ln{j}.
\end{align}}
Following Eq.~\ref{eq:numberOfGiantsDifferential}, but avoiding integration over $z$ and assuming narrow bins in both dimensions, we approximate
\begin{align}
    \lambda_i \approx n_\mathrm{GRG} V_i \cdot f_{L_\mathrm{p}\ \vert\ L_\mathrm{p} \geq l_\mathrm{p,GRG}}(l_{\mathrm{p},i}) \Delta l_\mathrm{p} \cdot p_\mathrm{obs}(l_{\mathrm{p},i},z_i).
\end{align}
The volume in which the giants of bin $i$ fall, $V_i$, is
\begin{align}
    V_i = \Omega r^2(z_i)\Delta r_i,\ \text{with}\ \Delta r_i = \frac{c}{H_0}\frac{\Delta z}{E(z_i)}.
\end{align}
Appendix~\ref{app:likelihood} details a particularly efficient trick to compute the likelihood for a range of $n_\mathrm{GRG}$, whilst leaving the other parameters fixed.
By multiplying the likelihood function with a prior distribution, for which we chose the uniform distribution, we obtained a posterior distribution over $\vec{\theta}$ -- up to a constant.

\section{Data}
\label{sec:data}
We applied our automated radio--optical catalogue creation methods to all Stokes $I$ maps from LoTSS DR2 \citep{Shimwell2022}.\footnote{LoTSS DR2 is publicly available at \url{https://lofar-surveys.org/dr2_release.html}.}
The LoTSS DR2 observations cover the $120$--$168\ \mathrm{MHz}$ frequency range, have a $6''$ resolution, a median RMS sensitivity of $83\ \mathrm{\mu Jy\ beam}^{-1}$, and a flux density scale uncertainty of approximately $10\%$.
The observations are split into a region centred at 12h45m $+44\degree 30'$ and a region centred at 1h00m $+28\degree 00'$; both avoid the Galactic Plane.
These regions span $4,178$ and $1,457$ square degrees respectively, and together cover $27\%$ of the Northern Sky.
The observations consist of $841$ partly overlapping pointings with diameters of $4.0\degree$.
The vast majority of the pointings were observed for 8h, all within the May 2014--February 2020 time frame.

The ML pipeline presented in Sect.~\ref{sec:methods} does not only rely on LoTSS DR2 Stokes $I$ maps, but also on an infrared--optical source catalogue.
This catalogue contains the positions, magnitudes, and colours of unWISE \citep{Schlafly2019} infrared sources and of DESI Legacy Imaging Surveys DR9 \citep{Dey2019a} optical sources.

To discover as many giants as possible, we supplemented our ML pipeline's sample of GRG candidates with the GRG candidates from the value-added LoTSS DR2 catalogue \citep{Hardcastle2023}.
For angularly extended ($\phi > 15''$) radio components (and thus all giants), the radio source component association and most of the host galaxy identification for the value-added LoTSS DR2 catalogue were performed via a public project named `Radio Galaxy Zoo: LOFAR' on Zooniverse.
Zooniverse is an online citizen science platform for crowd-sourced visual inspection.\footnote{The Zooniverse website is \url{https://zooniverse.org}.}
We will refer to the value-added LoTSS DR2 catalogue as the `RGZ catalogue' and to the GRG candidates in that catalogue as the `RGZ GRG candidates'.
The detailed source component information provided by the RGZ catalogue allowed us to homogenise the angular length estimates of the ML pipeline GRG candidates and the RGZ GRG candidates (see Sect.~\ref{sec:angular}).
After visual confirmation, we supplemented our GRG sample with literature GRG samples (see Sect.~\ref{sec:literature-sample}).

\section{Methods}
\label{sec:methods}

\begin{figure*}\begin{center}
\includegraphics[width=0.9\textwidth]{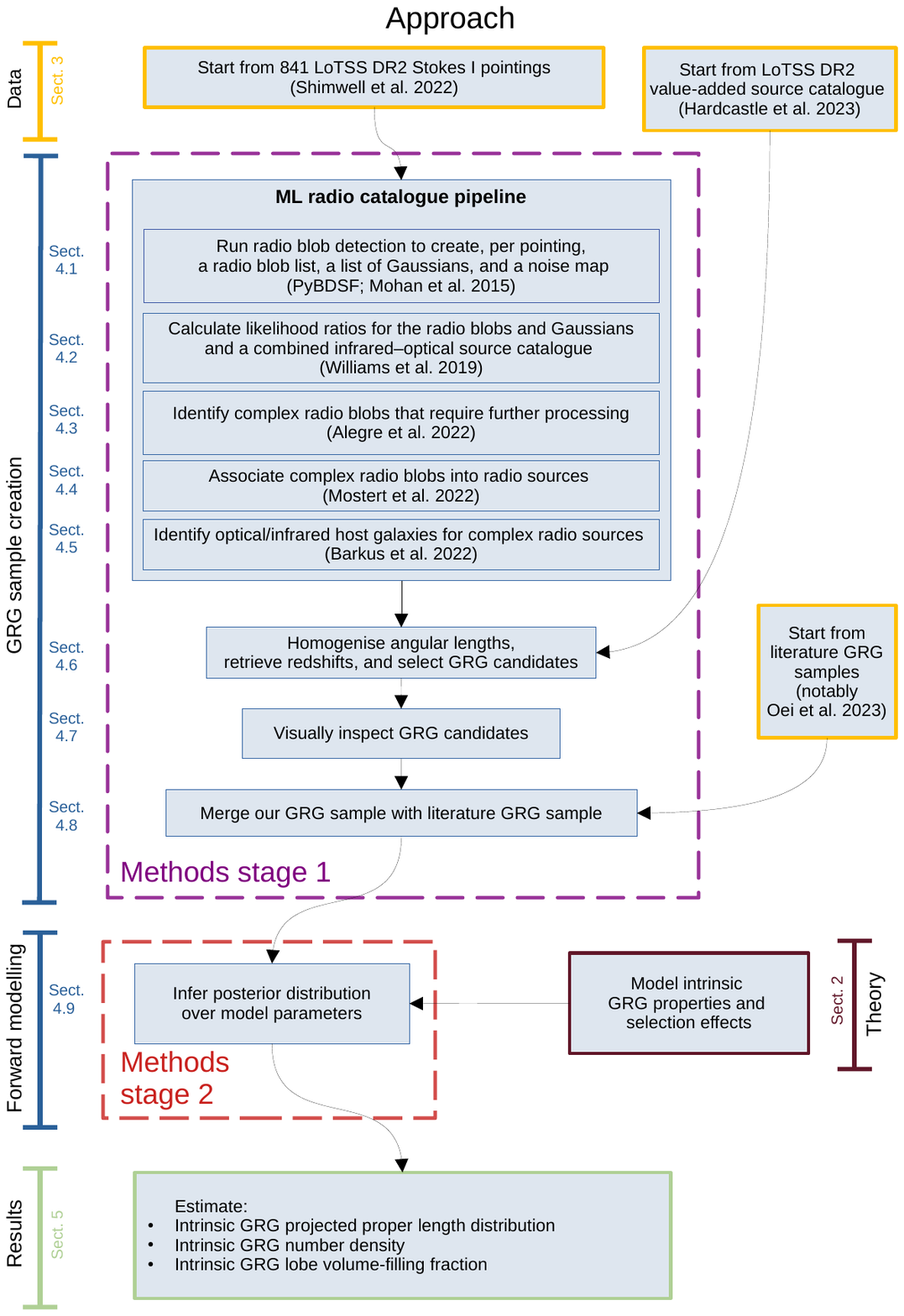}
\caption{
Overview of our approach, which consists of two stages.
In the first stage we built a GRG sample, and in the second stage we inferred the properties of the intrinsic GRG population using a forward model.
The brackets indicate the different parts of our approach and mention the sections containing the corresponding details.
} \label{fig:approach}
\end{center}\end{figure*}

To derive the projected length distribution, number density, and lobe VFF for the intrinsic population of giants, we followed a two-stage approach.
In the first stage, we gathered all giants that we detected in the LoTSS DR2 Stokes $I$ images using our automatic ML pipeline and added all other giants that we found in the RGZ catalogue.
We re-evaluated and homogenised the source size estimates over the combined GRG sample, and manually inspected the plausibility of the associated radio source components and optical/infrared host galaxy.
Finally, we merged this GRG sample with the other GRG samples from the literature.
In the second stage, we search for the most likely parameters for the forward model presented in Sect.~\ref{sec:theory} that describe the GRG observed projected proper length distribution and the selection effects of the merged GRG sample.
Figure~\ref{fig:approach} shows an overview of our approach.

\subsection{Detecting radio emission}
\label{sec:detection}
We started out with the publicly available calibrated LoTSS DR2 Stokes $I$ images \citep{Shimwell2022}.
For each of the $841$ pointings, we ran the PyBDSF radio blob detection software \citep{Mohan2015} using the same parameters as used in LoTSS DR2 \citep{Shimwell2022} -- notably, this means that we used a $5\sigma$ detection threshold.
Appendix~\ref{app:pybdsf} provides the full list of PyBDSF parameters and their values.

The output we generated consists of a list of radio blobs with their location and properties. 
PyBDSF can decompose each radio blob it detects, into one or more 2D Gaussians. For each radio blob, we also saved the corresponding list of Gaussians. 
These Gaussians function as a source model for each radio blob and will be used in later steps in the ML pipeline.\footnote{However, as we discuss in Sect. \ref{sec:angular} these source models are not always adequate for extended, well-resolved radio sources.}

\subsection{Calculating radio-to-optical/infrared likelihood ratios}
\label{sec:lr}
For radio sources, the location of the host galaxy on the sky is close to the flux density--weighted centre of the radio source.\footnote{In this context, `close' refers to angular distances comparable to the survey's resolution.}\footnote{The flux density--weighted centre of a radio component is the sum of the product of position and flux density for each pixel in the area where PyBDSF found significant emission, divided by the area's total flux density. This calculation is performed twice: once for determining the right ascension and once for determining the declination of the centroid. The flux density--weighted centre of a multi-component radio source is the flux density--weighted average of the components' individual flux density--weighted centres.}
The likelihood ratio method, which exploits this idea, quantifies the likelihood that a source in one observing band is the correct counterpart to a source in another observing band \citep[e.g.][]{Richter1975,deRuiter1977,Sutherland1992}.
\citet{Williams2019} used this method to cross-match the unresolved -- and some resolved -- radio sources of LoTSS DR1 to a combined catalogue of infrared and optical sources.
More specifically, the infrared sources came from AllWISE \citep[][]{Cutri2014}, whilst the optical sources came from the Panoramic Survey Telescope and Rapid Response System 1 \citep[Pan-STARRS1;][]{Chambers2016} DR1 $3\pi$ steradian survey.
The likelihood ratio function that \citet{Williams2019} used is a function of the angular distance between the flux density--weighted centre of the radio source and the flux density--weighted centre of the optical or infrared source, the magnitude of the optical or infrared source, and the colour of the optical or infrared source.
The likelihood ratio function also takes into account uncertainties in each of these three dependencies.

We adopted the same procedure as detailed by \citet{Williams2019} to cross-match our simple radio sources (where `simple' is to be understood as in Sect. \ref{sec:sorting}) to a combined catalogue of infrared and optical sources.
The infrared sources came from unWISE \citep{Schlafly2019}, and the optical sources were now taken from the DESI Legacy Imaging Surveys DR9 \citep{Dey2019a}, which boasts deeper imagery than Pan-STARRS1 DR1 used for LoTSS DR1.
The unWISE \citep{Schlafly2019} and DESI Legacy Imaging Surveys DR9 source catalogues are used for LoTSS DR2 cross-matching more generally \citep{Hardcastle2023}.
Per pointing, we applied the likelihood ratio method to the full list of radio blobs and to the full list of Gaussians.
For both the blobs and the Gaussians, we stored the identifier of the optical or infrared source that produced the highest likelihood ratio, alongside this highest likelihood ratio itself.

\subsection{Sorting radio emission with a gradient-boosting classifier}
\label{sec:sorting}
Most radio sources that consist of a single radio blob (mostly unresolved or barely resolved radio sources) can be cross-matched using the likelihood ratio method.
However, some resolved radio sources, and certainly most resolved giants (Sect.~\ref{sec:theoryAnglengths}), consist of multiple radio blobs as parametrised by PyBDSF, and therefore require radio blob association and cannot be cross-matched using the likelihood ratio alone.
To separate the simple from the complex radio blobs in LoTSS DR1, a considerable amount of visual inspection was applied \citep{Williams2019}.
For LoTSS DR2, \citet{Alegre2022} trained a gradient-boosting classifier \citep[GBC;][]{breiman1997arcing, friedman2001greedy} to classify radio blobs as either `simple' or `complex' based on the properties of the radio blobs, the properties of the Gaussians fitted to these blobs, the likelihood ratios for each, and the distance to and properties of the nearest neighbours.

We adopted the procedure of \citet{Alegre2022} and use their trained GBC to separate the simple radio blobs from those that require radio component association beyond PyBDSF's capabilities and/or optical host identification beyond the scope of the likelihood ratio method of \citet{Sutherland1992}.
We expect most giants to fall in the latter case.

\subsection{Associating radio emission into radio sources}
\label{sec:association}
We proceeded with automatic radio source component association for the complex radio blobs.
Following the procedure laid out by \citet{Mostert2022}, for each of these radio blobs, we created a $300'' \times 300''$ LoTSS DR2 image cutout centred on the radio blob.
Next, a fast region-based convolutional neural network \citep[Fast R-CNN;][]{girshick2015fast}, adapted and trained for this purpose by \citet{Mostert2022}, was applied to these cutouts to predict which (if any) other radio blobs -- whether they be complex or simple -- form a single physical structure with the central radio blob.
For example, the two lobes of an RG, each represented by a radio blob, might be associated together to form a single physical radio source.
Due to the fixed $300'' \times 300''$ image size for which the Fast R-CNN was trained, we expect most radio sources that are associated in our pipeline to have an angular length $\phi < 424''$.\footnote{If predicted associations from neighbouring cutouts have an overlapping radio blob, the associations will be merged. For example: in cutout 1 lobe A and core B are associated and in cutout 2 core B and lobe C are associated, then the set (lobe A, core B, and lobe C) will enter the catalogue as a single radio source, thereby creating the possibility of detecting radio sources with angular length $\phi > 424''$.}

The result is a radio source catalogue in which some of the radio blobs have been merged, and a component catalogue that lists for each radio blob to which radio source it belongs.
The radio and the component catalogue were completed by appending to them the remaining list of simple radio blobs.

\subsection{Identifying host galaxies in the optical and infrared}
\label{sec:identification}
\citet{Barkus2022} created a method for identifying the optical or infrared host of an extended radio source.
The method described by \citet{Barkus2022} takes the radio morphology into account by drawing a ridgeline along the regions of high flux density. 
The method continues with the application of the likelihood ratio method to quantify which pairs of host galaxy candidates and radio sources are a plausible match.
The likelihood ratio $LR$ used in this context follows Eq.~1 of \citet{Sutherland1992}, with the slight simplification of having the latter's dependence on two angular offsets replaced by a dependence on a \emph{radial} angular offset only:
\begin{align}
\label{eq:LR}
    LR = \frac{q(m,c) f(r)}{n(m,c)},
\end{align}
where $q(m,c)$ is a prior on the magnitude $m$ and colour $c$ of the optical host, $f(r)$ is a function of the angular offset between the optical centroid and the radio centroid, and $n(m,c)$ normalises for the number density of optical sources with a certain magnitude $m$ and colour $c$ in the catalogue used for the cross-matching.

To adapt the likelihood ratio for use in the case of extended radio sources, \citet{Barkus2022} implemented the different components of the ratio as follows. 
For $n(m,c)$, \citet{Barkus2022} estimated the probability density over $m$ and $c$ for a distribution of $50,000$ randomly sampled sources from a combined Pan-STARRS--AllWISE catalogue in the region of the sky that overlapped with LoTSS DR1. 
For $q(m,c)$, they estimated the probability over $m$ and $c$ for sources from the combined Pan-STARRS--AllWISE catalogue that were manually selected to be the most likely optical/near-infrared host for a sample of $950$ radio sources with angular length $\phi > 15''$.
For both $n(m,c)$ and $q(m,c)$, the AllWISE W1 magnitudes were used for $m$, the Pan-STARRS i-band magnitudes minus the AllWISE W1 magnitudes were used for colour $c$, and the PDF was formed using a 2D kernel density estimator \citep[KDE; e.g.][]{Pedregosa2011} with a Gaussian kernel and a bandwidth of $0.2$.
For extended asymmetric or bent radio galaxies, the optical host is not likely to be found at the radio centroid.
Therefore, \citet{Barkus2022} proposed that $f(r)$ should be a function of both the distance between the radio centroid and the optical source $r_\mathrm{opt,centroid}$ and the smallest distance between the optical source and a ridgeline fitted to the radio source $r_\mathrm{opt,ridge}$.
Specifically,
\begin{equation}
f(r)=f_\mathrm{ridge}(r_\mathrm{opt,ridge})\cdot f_\mathrm{centroid}(r_\mathrm{opt,centroid}),
\end{equation}
with
\begin{equation} \label{eq:fridge}
f_\mathrm{ridge}(r_\mathrm{opt,ridge})=\frac{1}{2 \pi \sigma_r^2} e^{\frac{-r_\mathrm{opt,ridge}^2}{2\sigma_r^2}},
\end{equation}
and
\begin{equation} \label{eq:fcentroid}
f_\mathrm{centroid}(r_\mathrm{opt,centroid})=\frac{1}{2 \pi \sigma_\mathrm{c}^2} e^{\frac{-r_\mathrm{opt,centroid}^2}{2\sigma_\mathrm{c}^2}},
\end{equation}
where $\sigma_r^2=\sigma_\mathrm{opt}^2 + \sigma_\mathrm{radio}^2 + \sigma_\mathrm{astr}^2$.
We fixed the astrometric uncertainty $\sigma_\mathrm{astr} = 0.2''$.
The optical position uncertainties $\sigma_\mathrm{opt}$ are taken from the optical catalogue (generally ${\sim}0.1''$), the radio position uncertainty $\sigma_\mathrm{radio}$ is fixed to $3''$, and the uncertainty in the centroid position $\sigma_\mathrm{c}$ is empirically estimated at $0.2$ times the length of the considered radio source.
For the $30$ optical sources closest to the radio ridgeline, \citet{Barkus2022} calculated the $LR$ and considered the source with the highest $LR$ to be the most likely host galaxy.

We used the method of \citet{Barkus2022} but made three minor adaptations.
First, we introduce explicit regularisation for $q(m,c)$ and $n(m,c)$.
As the PDF estimates for $q(m,c)$ and $n(m,c)$ are 2D KDEs over sampled $(m,c)$-distributions, the parts of the $(m,c)$--parameter space that are sparsely sampled can lead to probabilities that are effectively zero when the realistic theoretical probability should be small but non-zero.
Through the $q/n$-fraction in Eq. \ref{eq:LR}, the resulting values of $LR$ in the sparsely sampled parts of the $(m,c)$--parameter space blow up to unrealistic large values or collapse to almost 0 (see Fig. \ref{fig:unregularised}). In practice, these unsampled parts of parameter space are almost never visited by new sources for which we calculate $LR$. 
Even so, we add a constant factor to the KDE estimate of $q$ and $n$ to get more robust $LR$ values (see Fig. \ref{fig:regularised}) and to express the model uncertainties in our functions of $q$ and $n$.
Using 10-fold cross-validation, we empirically select the bandwidths for the KDEs leading to $q$ and $n$ to be $0.4$.
Second, we propose an alternate form of $f(r)$.
For giants, $f(r)$ is rarely dominated by errors in the position of the optical source or that of the radio source.
As $r_\mathrm{opt,centroid}$ and $r_\mathrm{opt,ridge}$ are slightly correlated, multiplication of $f_\mathrm{ridge}(r_\mathrm{opt,ridge})$ and $f_\mathrm{centroid}(r_\mathrm{opt,centroid})$ under-estimates the chance of low values of $r_\mathrm{opt,centroid}$ or $r_\mathrm{opt,ridge}$. 
Therefore, we combine $r_\mathrm{opt,centroid}$ and $r_\mathrm{opt,ridge}$ into a single parameter $r_\mathrm{mean}$ that is the mean of the two distance parameters.
Furthermore, we observe that the empirical distributions of $r_\mathrm{opt,centroid}$, $r_\mathrm{opt,ridge}$ and $r_\mathrm{mean}$ for a sample of radio sources with angular length $>1'$ $A_\mathrm{radio,opt}$ for which optical counterparts were determined via visual inspection do not follow a normal distribution as assumed by \citet{Barkus2022} but rather a lognormal distribution (see Fig. \ref{fig:f(r)}).
Instead of estimating the values of the different error components (astrometric error, error in optical position, error in radio position) we use the empirical values of the distribution of $f(r)$ for the sources in $A_\mathrm{radio,opt}$; see Appendix \ref{app:ridge} for details.
Third, we replaced the Pan-STARRS1 DR1 catalogue (from which colour $c$ was derived) with the DESI Legacy Imaging Surveys DR9 catalogue, as the latter goes up to an $i$-band magnitude of 24.

We applied the modified ridgeline method to all radio sources in our pipeline catalogue with angular lengths larger than $1'$ and brighter than $10\ \mathrm{mJy}$.
We limit the ridgeline procedure to these sources to save time, as the procedure takes multiple seconds per radio source.

After detecting the host galaxies of our radio sources, we checked for spectroscopic redshifts from SDSS \citep[VizieR catalogue \texttt{V/147/sdss12};][]{Ahn2012}, or if not available, for photometric redshifts from DESI \citep[VizieR catalogue \texttt{VII/292};][]{Duncan2022}. 
The SDSS catalogue also provides velocity dispersions and a quasar flag.
The DESI catalogue includes a flag (\textsc{fclean}) that indicates whether the optical source used in photometric redshift estimation is free from blending and image artefacts.
The catalogue also includes a column (\textsc{pstar}) that estimates how likely it is that the optical source is a star based on its colours.
In both the ML pipeline and RGZ catalogues, we only retained sources for which \textsc{fclean} $=1$ and \textsc{pstar} $\leq 0.2$.

\subsection{Reassessing angular source lengths}
\label{sec:angular}
\begin{figure}
\centering
\includegraphics[width=\columnwidth]{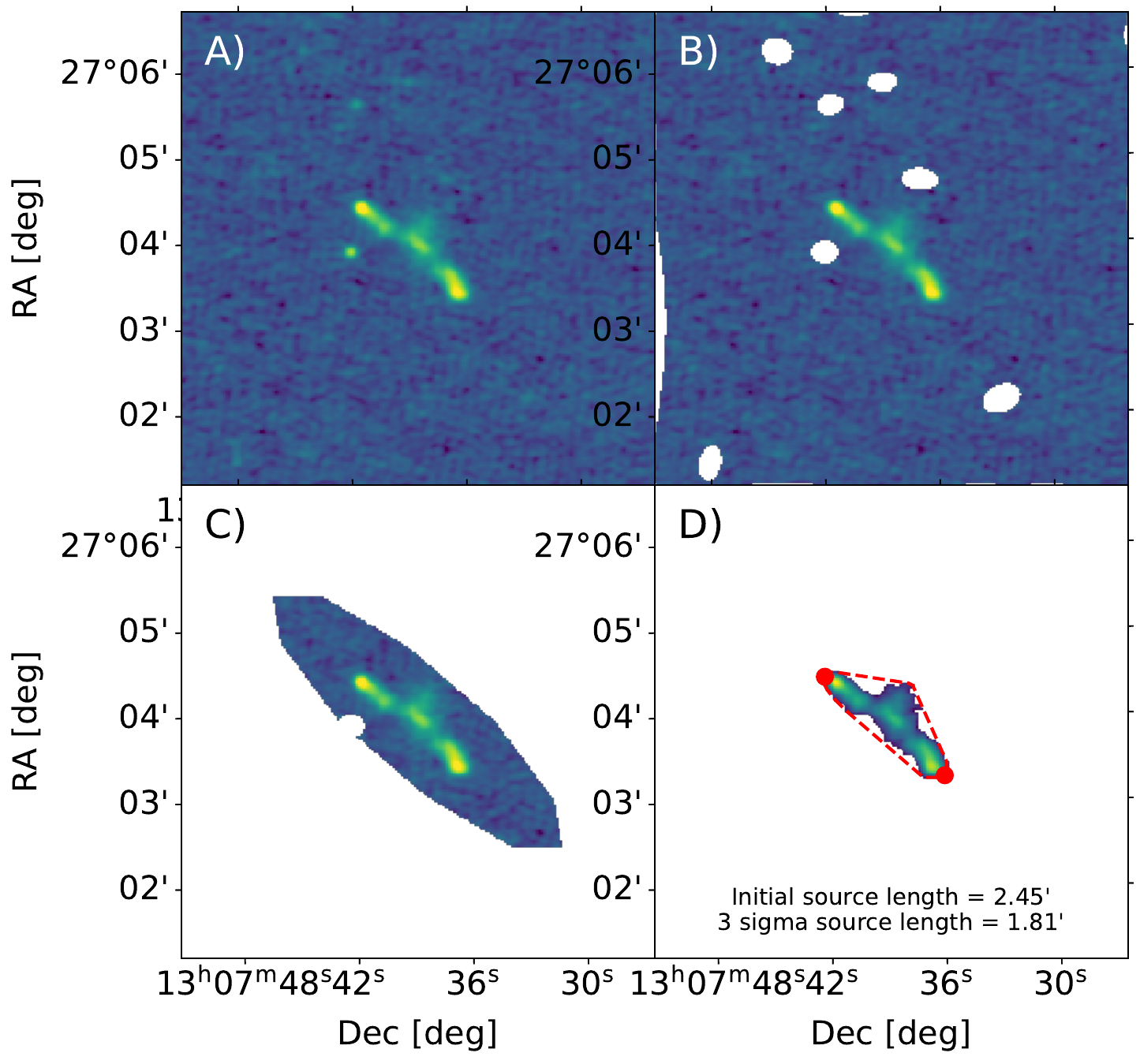}
\caption{
Summary of the angular length re-evaluation for radio source ILTJ130738.79+270355.1.
Panels A--D show the initial cutout, the removal of neighbouring sources, the masking of emission outside a convex hull based on the old angular length, and the emission that is left after masking all emission below thrice the local noise $\sigma$.
The red line segments delineate the convex hull of the left-over emission, and the red points indicate the furthest removed points in this convex hull.
The great-circle distance between these points is the $3\sigma$ angular length.
}
\label{fig:reassess}
\end{figure}

Next, we proceeded to reassess the angular source lengths, for both the radio catalogue created using the ML pipeline and the \texttt{Composite\_Size} column reported by the RGZ catalogue described by \citet{Hardcastle2023}.
The angular source lengths in the ML catalogue and the \texttt{Composite\_Size} column in the RGZ catalogue are the full width at half maximum (FWHM) of the combined Gaussian components that make up a source, if the source is only composed of a single radio blob.
If the radio source is composed of multiple radio blobs, the reported size is the distance between the two furthest removed points on a convex hull that encloses the FWHMs of the blobs that make up the radio source.
However, in the literature, the length of a GRG is often reported to be the maximum distance between the signal of a radio source that exceeds three times the image noise $\sigma$.

To get the $3\sigma$ angular lengths, we applied five steps to all sources in both catalogues with a reported angular length $\phi > 1'$.
First, we created a square image cutout with a width and height equal to $1.5$ times the old angular source length.
Second, we mask all neighbouring radio emission. 
Third, we mask all emission outside an ellipse with a major axis that is the old source length, and a minor axis that is 1.1 times the old source width or a quarter of the old source length if that value is bigger.
These numbers are a result of the observation that, with respect to the $3\sigma$ angular lengths, the old lengths were almost always significantly overestimated, while the source width tended to be underestimated. 
Fourth, we mask all remaining emission that is below three times the local noise. 
Fifth, we fitted a convex hull around the remaining emission and determined the distance between the points on this convex hull that were farthest apart. 
See Fig.~\ref{fig:reassess} for an illustrative example.

The entire process from source detection (Sect. \ref{sec:detection}) to source list with optical identifications and updated angular lengths (this subsection) took roughly half an hour to one hour per LoTSS DR2 pointing, depending on the detected number of sources. %
Each pointing can be processed independently, which allowed us to spread the processing of all $841$ LoTSS DR2 pointings over 5 nodes of a heterogeneous computer cluster with 80 physical CPU cores in total for three to four days.

Finally, for both the ML pipeline and RGZ catalogues, we calculate the projected proper lengths using the $3\sigma$ angular lengths and the redshift estimates corresponding to each source, and discard all sources that do not meet the $l_\mathrm{p} \geq l_\mathrm{p,GRG}$ criterion.
For the ML pipeline catalogue, we discarded all internally duplicate GRG candidates using a $1'$ cone search.
The RGZ catalogue did not contain any internal duplicates.
That left us with $7,001$ GRG candidates in the ML pipeline catalogue and $7,044$ GRG candidates in the RGZ catalogue.

\subsection{Inspecting GRG candidates visually}
\label{sec:verification}
The following step we took in the creation of our GRG sample was a manual visual inspection of all our GRG candidates.
For the RGZ GRG candidates, as described by \citet{Hardcastle2023}, at least five different volunteers already inspected the radio and corresponding optical emission, and in most cases the candidates identified in this way were reinspected by a professional astronomer.
The purpose of our manual visual inspection was therefore to exclude only those sources where either the radio component association or the host identification was obviously incorrect.
For each GRG candidate, a single expert looked at a panel showing the candidate with its neighbouring sources masked and most neighbouring emission masked (akin to panel C in Fig.~\ref{fig:reassess}) and a panel showing the candidate in its wider context (akin to panel A in Fig.~\ref{fig:reassess}); additionally, the location of the optical host was indicated.
We sorted the candidates into three categories: candidates that looked reasonable, candidates that clearly missed (or included too many) significant radio components, and candidates that showed a very unlikely host galaxy location.
For the ML pipeline GRG candidates, we initially followed the same procedure as for the RGZ GRG candidates.
To speed up the visual inspection, we skipped the $4,272$ ML pipeline GRG candidates that were verified RGZ giants.
After inspecting the ML pipeline GRG candidates once, we subjected all that were not rejected to a second round of visual inspection.
The second round was aided by inspecting LoTSS DR2 radio contours over a Legacy Survey DR9 $(g,r,z)$ image cube, where sources from the combined optical--infrared catalogue within the field of view were highlighted.

For the RGZ catalogue, we judged $6,550$ ($93\%$) GRG candidates to be without issues, $389$ ($6\%$) to have radio component issues, and $105$ ($1\%$) to have been assigned an unlikely host galaxy.
For the $5,864$ (unique) ML pipeline GRG candidates, we judged $2,722$ ($47\%$) candidates to be without issues, $1,963$ ($33\%$) to have radio component issues, and $1,179$ ($20\%$) to have been assigned an unlikely host galaxy.
Radio component association issues for the ML-identified candidates occur because the association method leverages an object detection neural network with rectangular bounding boxes to capture the radio components \citep{Mostert2022}.
The large extent of these sources causes many unrelated (fore- and background) radio sources to appear in the rectangular bounding box that encompasses the candidate, increasing the likelihood of erroneous component associations.
Future ML radio association methods should consider using instance segmentation instead of object detection with a rectangular bounding box.\footnote{For the distinction between object detection and instance segmentation, see \citet{lakshmanan2021practical}.}
Of the $6,550$ RGZ giants, $5,647$ do not appear in previous literature and are thus new discoveries.
Of the $2,722$ ML pipeline giants that are not RGZ giants, $2,597$ are new discoveries.

\begin{figure*}
    \centering
    \begin{subfigure}{\columnwidth}
    \includegraphics[width=\columnwidth]{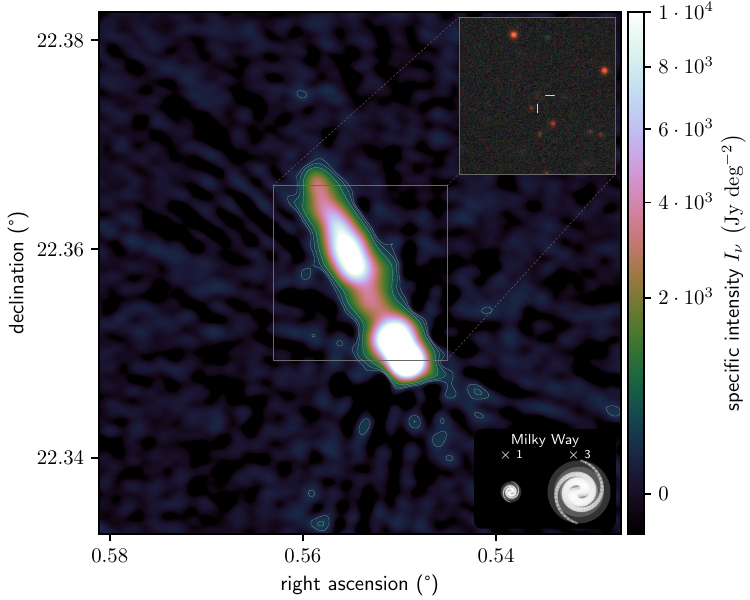}
    \end{subfigure}
    \begin{subfigure}{\columnwidth}
    \includegraphics[width=\columnwidth]{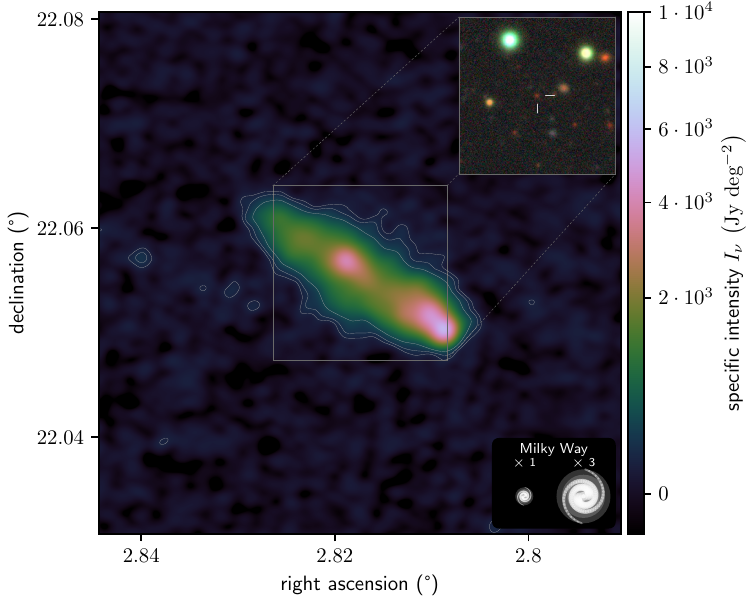}
    \end{subfigure}
    \begin{subfigure}{\columnwidth}
    \includegraphics[width=\columnwidth]{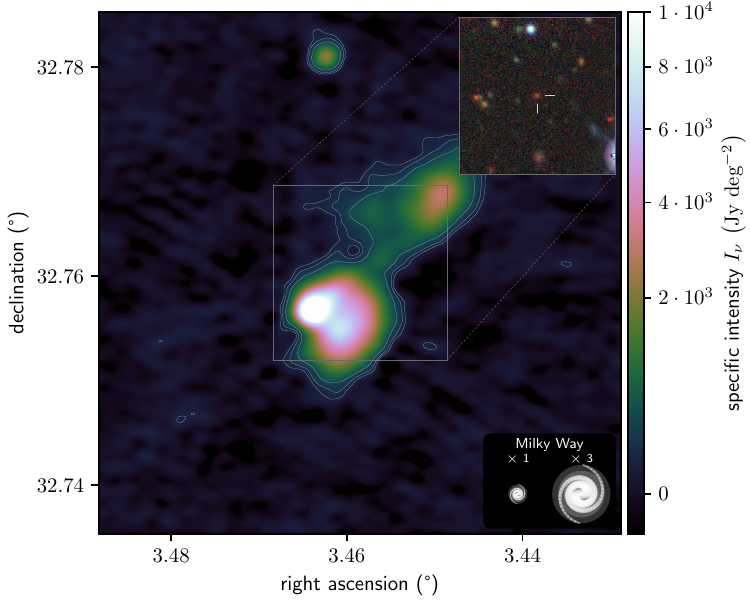}
    \end{subfigure}
    \begin{subfigure}{\columnwidth}
    \includegraphics[width=\columnwidth]{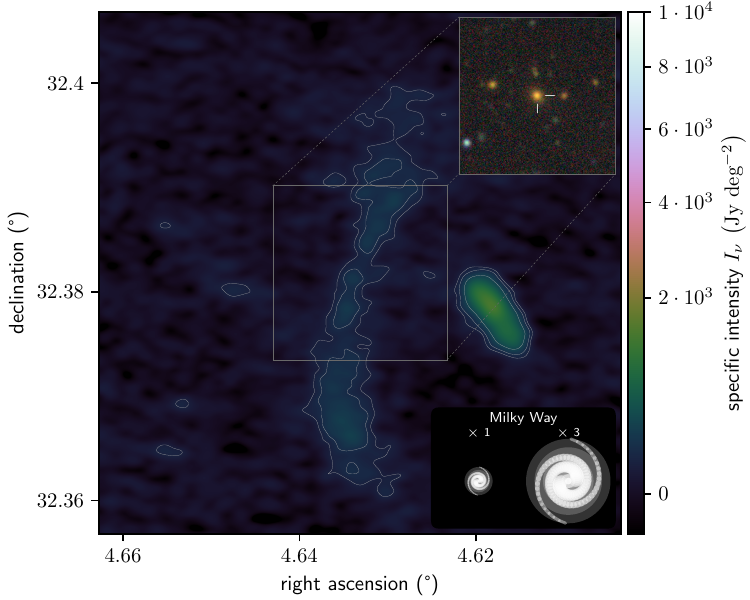}
    \end{subfigure}
    \begin{subfigure}{\columnwidth}
    \includegraphics[width=\columnwidth]{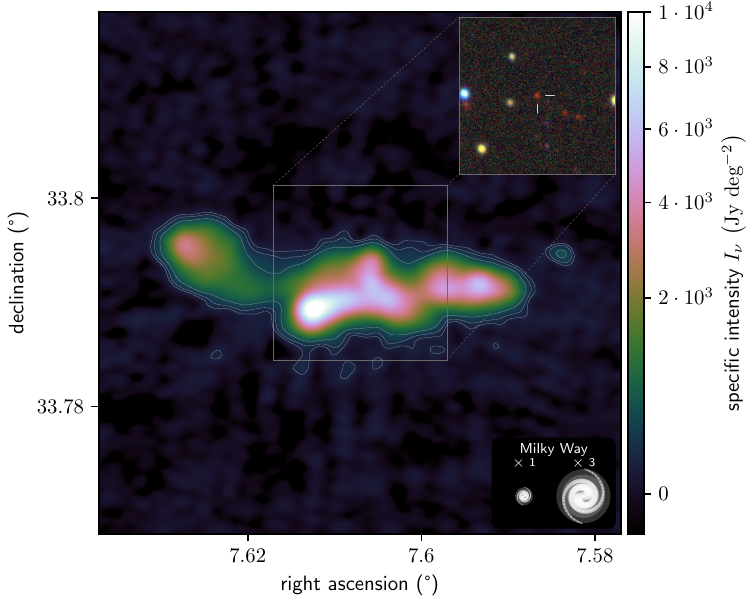}
    \end{subfigure}
    \begin{subfigure}{\columnwidth}
    \includegraphics[width=\columnwidth]{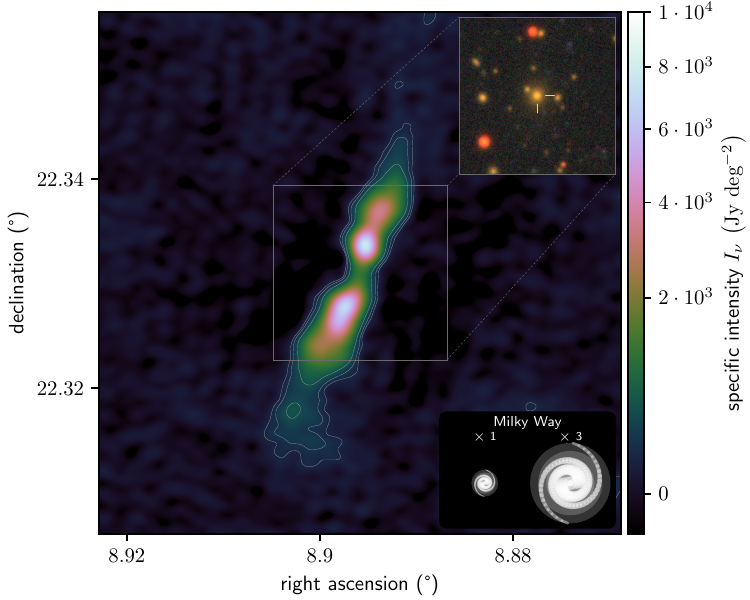}
    \end{subfigure}
    \caption{
    LoTSS DR2 cutouts at central observing frequency $\nu_\mathrm{obs} = 144\ \mathrm{MHz}$ and resolution $\theta_\mathrm{FWHM} = 6''$, centred around the hosts of newly discovered giants.
    Each cutout covers a solid angle of $3' \times 3'$.
    Contours signify 3, 5, and 10 sigma-clipped standard deviations above the sigma-clipped median.
    For scale, we show the stellar Milky Way disk (with a diameter of $50\ \mathrm{kpc}$) generated using the \citet{Ringermacher2009} formula, alongside a 3 times inflated version.
    Each DESI Legacy Imaging Surveys DR9 $(g,r,z)$ inset shows the central $1' \times 1'$ square region.
    As all giants obey $\phi \geq 1.3'$, they must -- if not oriented along one of the square's diagonals -- necessarily protrude from this region.
    Rowwise from left to right, from top to bottom, these giants are ILTJ000212.45+222116.2, ILTJ001115.77+220316.6, ILTJ001350.25+324530.8, ILTJ001831.84+322247.7, ILTJ003025.90+334729.2, and ILTJ003534.45+221937.8.
    }
    \label{fig:discoveries1}
\end{figure*}

\begin{figure*}
    \centering
    \begin{subfigure}{\columnwidth}
    \includegraphics[width=\columnwidth]{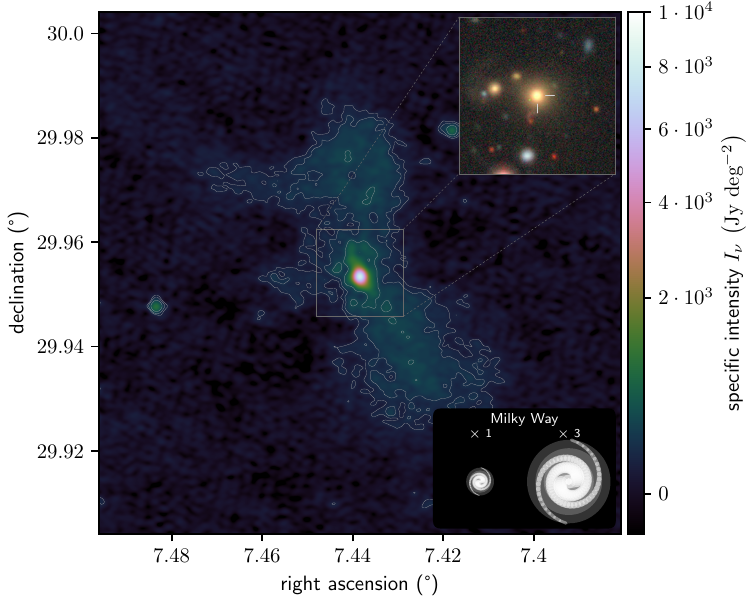}
    \end{subfigure}
    \begin{subfigure}{\columnwidth}
    \includegraphics[width=\columnwidth]{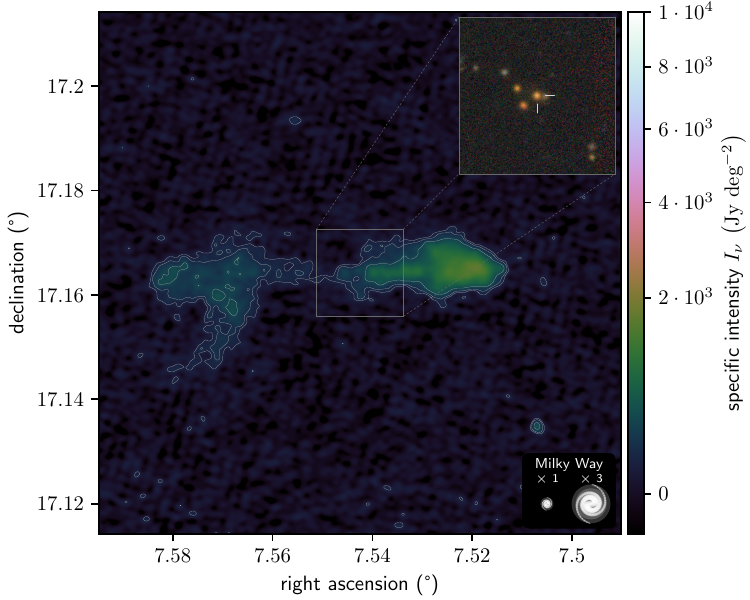}
    \end{subfigure}
    \begin{subfigure}{\columnwidth}
    \includegraphics[width=\columnwidth]{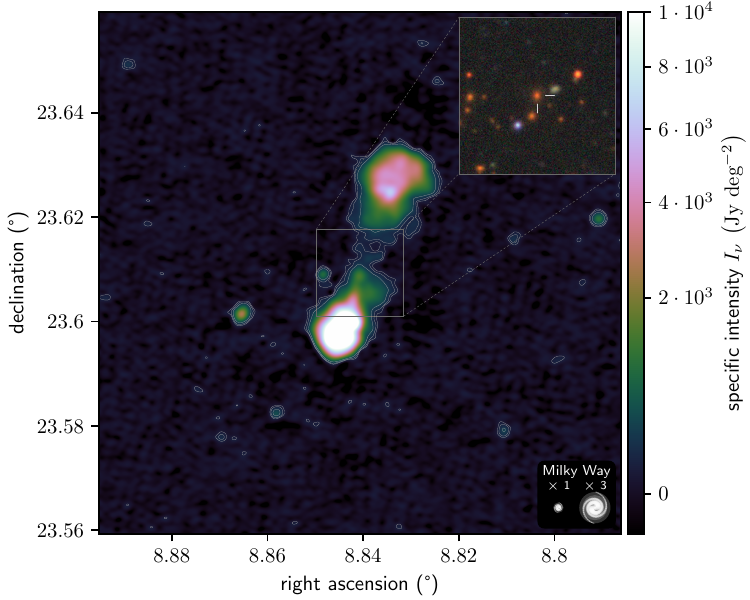}
    \end{subfigure}
    \begin{subfigure}{\columnwidth}
    \includegraphics[width=\columnwidth]{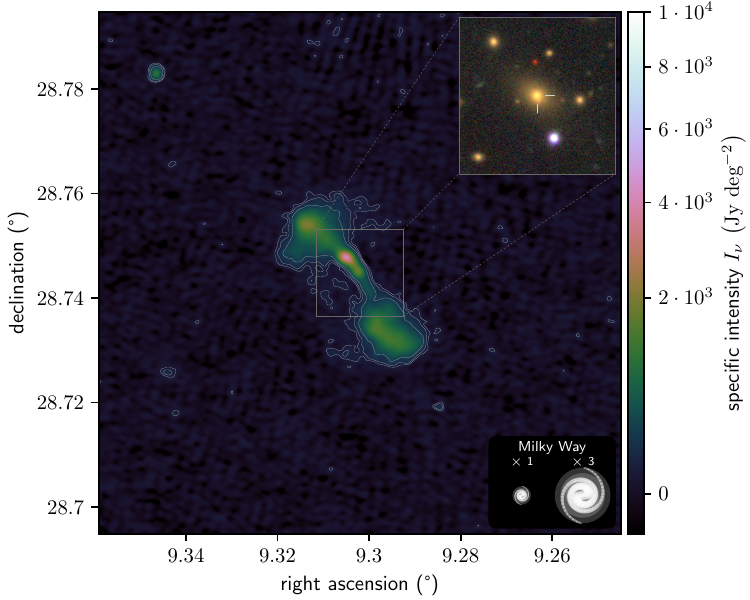}
    \end{subfigure}
    \begin{subfigure}{\columnwidth}
    \includegraphics[width=\columnwidth]{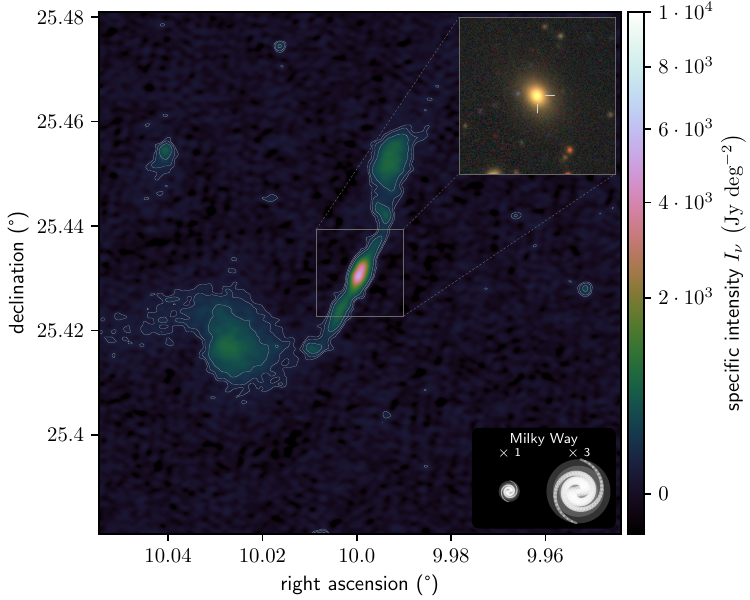}
    \end{subfigure}
    \begin{subfigure}{\columnwidth}
    \includegraphics[width=\columnwidth]{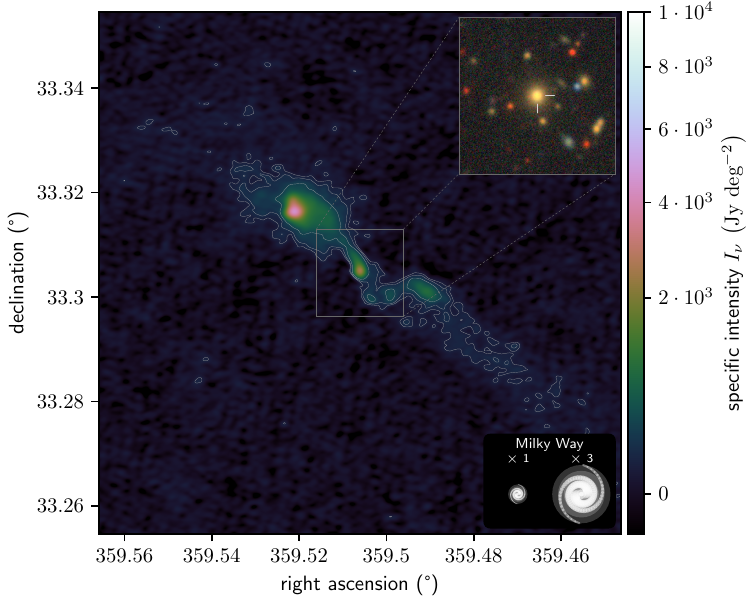}
    \end{subfigure}
    \caption{
    LoTSS DR2 cutouts at central observing frequency $\nu_\mathrm{obs} = 144\ \mathrm{MHz}$ and resolution $\theta_\mathrm{FWHM} = 6''$, centred around the hosts of newly discovered giants.
    Each cutout covers a solid angle of $6' \times 6'$.
    Contours signify 3, 5, and 10 sigma-clipped standard deviations above the sigma-clipped median.
    For scale, we show the stellar Milky Way disk (with a diameter of $50\ \mathrm{kpc}$) generated using the \citet{Ringermacher2009} formula, alongside a 3 times inflated version.
    Each DESI Legacy Imaging Surveys DR9 $(g,r,z)$ inset shows the central $1' \times 1'$ square region.
    As all giants obey $\phi \geq 1.3'$, they must -- if not oriented along one of the square's diagonals -- necessarily protrude from this region.
    Rowwise from left to right, from top to bottom, these giants are ILTJ002943.72+295700.3, ILTJ003010.58+170948.6, ILTJ003521.87+233625.9, ILTJ003712.91+284436.8, ILTJ004002.30+252550.9, and ILTJ235802.49+331838.5.
    }
    \label{fig:discoveries2}
\end{figure*}

Qualitatively, from the visual inspection, we noticed that the verified ML pipeline GRG sample contained more symmetric giants with colinear jets, while the verified RGZ GRG sample contained more giants with complex, bent structures indicative of interaction with the IGM.
The ML pipeline did also detect giants with complex structures, but was often unable to fully separate them from all neighbouring unrelated emission.
An in-depth comparison between the ML pipeline and RGZ GRG samples is beyond the scope of this work.
Figures~\ref{fig:discoveries1} and \ref{fig:discoveries2} each show six examples of previously unknown giants found through our ML-based approach.
Through cutouts covering $3' \times 3'$, Fig.~\ref{fig:discoveries1} shows angularly compact giants; through cutouts covering $6' \times 6'$, Fig.~\ref{fig:discoveries2} shows more angularly extended specimina.

\begin{figure*}
\includegraphics[width=\textwidth]{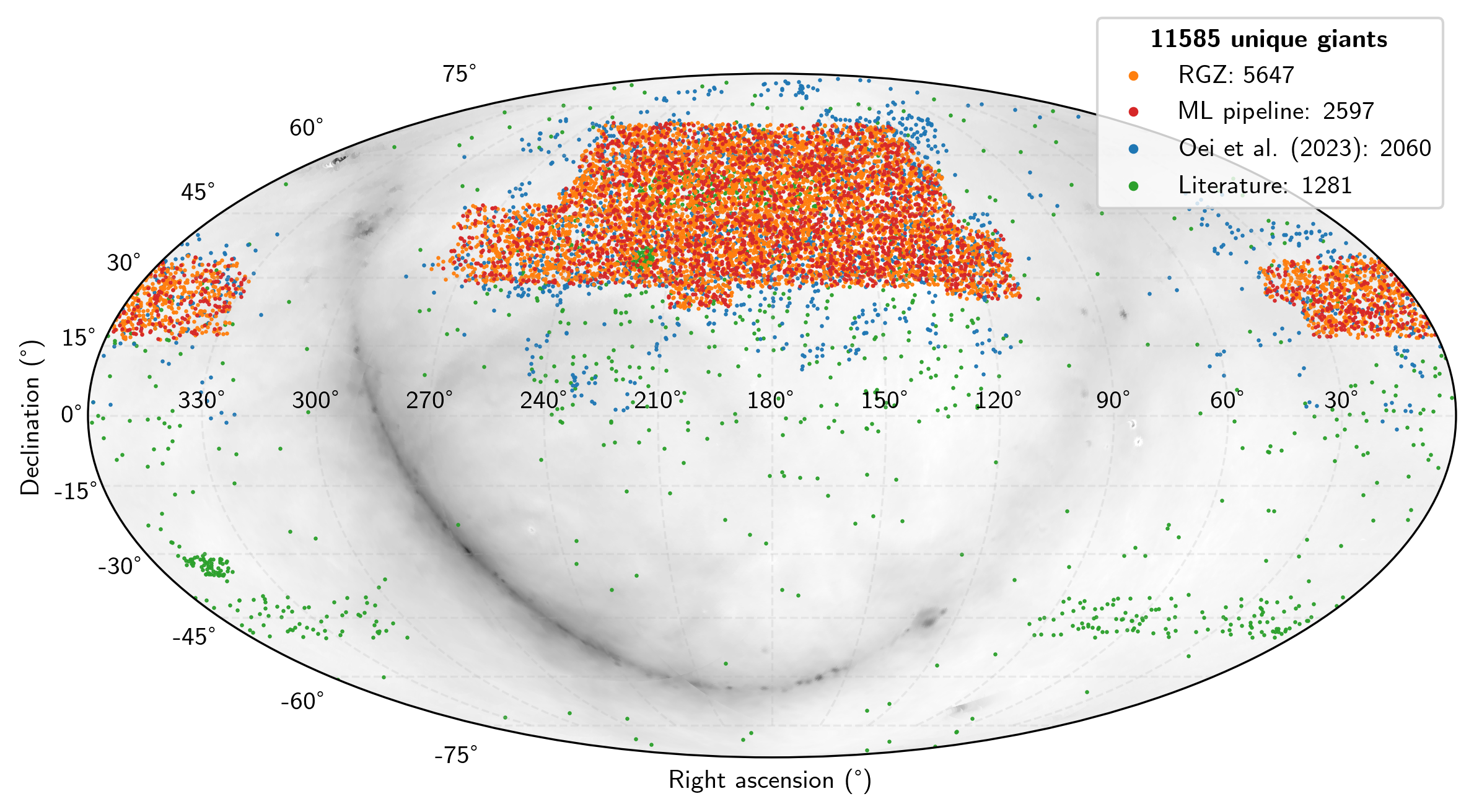}
\caption{With $11,585$ unique giants, we present the largest catalogue of large-scale galactic feedback to the Cosmic Web.
The RGZ (orange) and ML pipeline (red) samples are strictly confined to the LoTSS DR2 area, while the sample by \citet{Oei2023Distribution} extends to yet-to-be-released LoTSS pointings processed with the DR2 pipeline.
}
\label{fig:skyMap}
\end{figure*}

\subsection{Merging RGZ, ML pipeline, and literature samples}
\label{sec:literature-sample}
To complete our GRG sample, we iteratively added giants from the literature, going from the newest to the oldest publication.
This approach follows from the assumption that newer publications are generally based on more sensitive and higher-resolution observations, leading to more accurate angular length measurements.
In an effort to avoid having duplicate giants in the final sample, we only added giants when their host galaxies were more than $10''$ away from all host galaxies of already aggregated giants.

The joint RGZ--ML pipeline sample contains $9,272$ giants.
We added $1,471$ out of the $2,193$ giants presented by \citet{Oei2023Distribution}, 
$43$ out of the $69$ giants presented by \citet{Simonte2022}, 
$62$ out of the $62$ giants presented by \citet{Gurkan2022}, 
$165$ out of the $263$ giants presented by \citet{Mahato2022}, 
$178$ out of the $178$ giants presented by \citet{Andernach2021}, 
$0$ out of the $1$ giants presented by \citet{Masini2021}, 
$2$ out of the $2$ giants presented by \citet{Delhaize2021}, 
$1$ out of the $2$ giants presented by \citet{Bassani2021}, 
$1$ out of the $4$ giants presented by \citet{Tang2020}, 
$390$ out of the $694$ giants presented by \citet{Dabhade2020March}, and
$0$ out of the $6$ giants presented by \citet{Ishwara2020}.
These additions result in a final catalogue with $11,585$ unique giants.
This is the first catalogue of giants to contain more than $10^4$ specimina.

\begin{figure}
\includegraphics[width=\columnwidth]{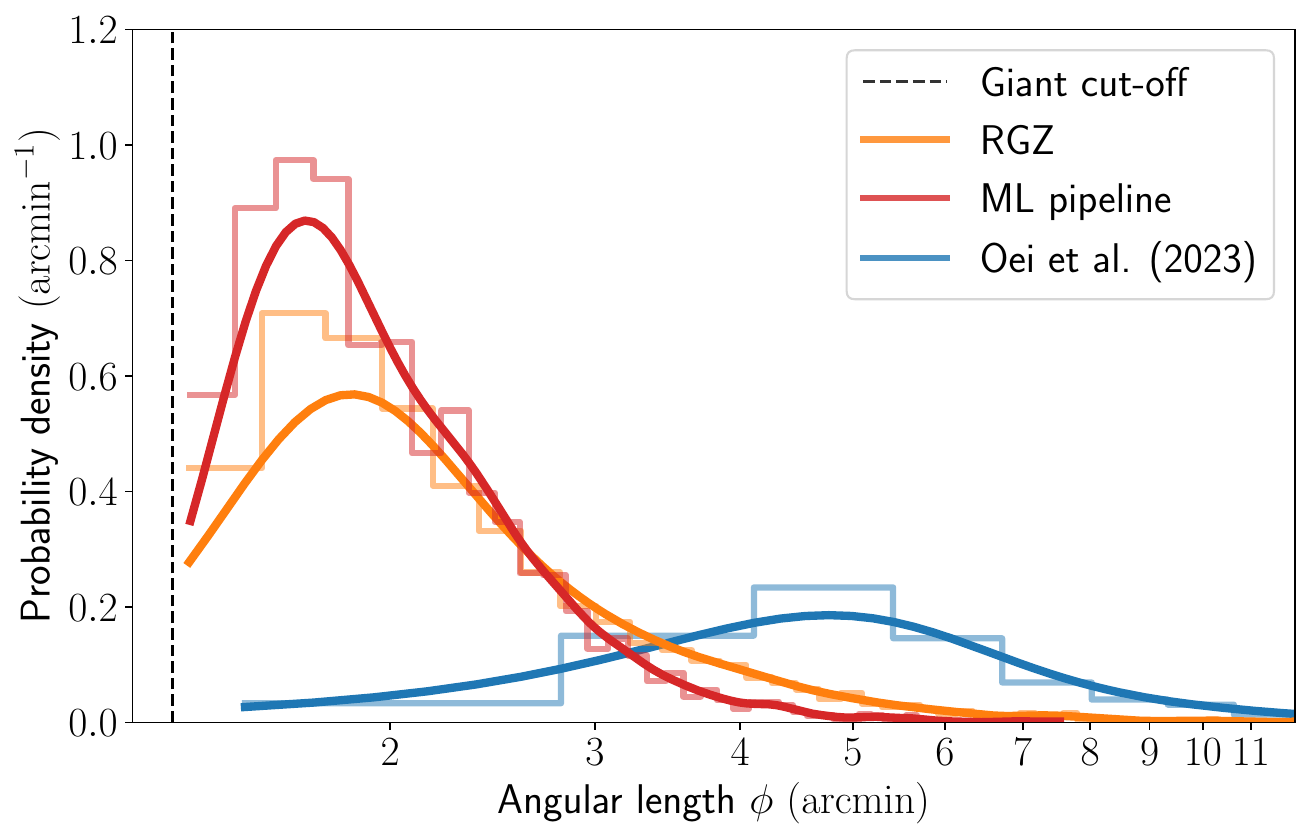}
\caption{
Observed distributions of angular length $\phi$, showing that our three LoTSS DR2 search methods target different ranges of $\phi$.
The largest angular lengths detected by \citet{Oei2023Distribution}, RGZ, and the ML pipeline are $132'$, $43'$, and $8'$ respectively, but we limit the horizontal axis to $12'$ for interpretability.
The vertical line marks the minimum angular length that giants can attain: $\phi_\mathrm{GRG}(l_\mathrm{p,GRG} = 0.7\ \mathrm{Mpc}) = 1.3'$.
} \label{fig:angular_length_hists}
\end{figure}

Figure~\ref{fig:skyMap} shows a Mollweide view of the sky with the locations of both the newly confirmed giants and the giants from the literature.
Almost all known giants stay clear of the Galactic Plane, where radio emission from the Milky Way -- of which we show the specific intensity function at $\nu_\mathrm{obs} = 150\ \mathrm{MHz}$ in greyscale \citep{Zheng2017} -- makes calibration and imaging harder.
In addition, optical host identification is much harder near the Galactic Plane.
The default field of view set-up of both our ML pipeline (Sect. \ref{sec:association}) and of RGZ favours the discovery of giants with angular lengths of a few arcminutes at most.
By contrast, the GRG search campaign of \citet{Oei2023Distribution} featured a `fuzzy' ${\sim}5'$ lower threshold to allow for an exhaustive manual search with an interactive and dynamic field of view \citep[using Aladin;][]{Bonnarel2000}.
Figure~\ref{fig:angular_length_hists} demonstrates that these design choices lead to GRG samples with markedly different angular length distributions.

\begin{figure}
\includegraphics[width=\columnwidth]{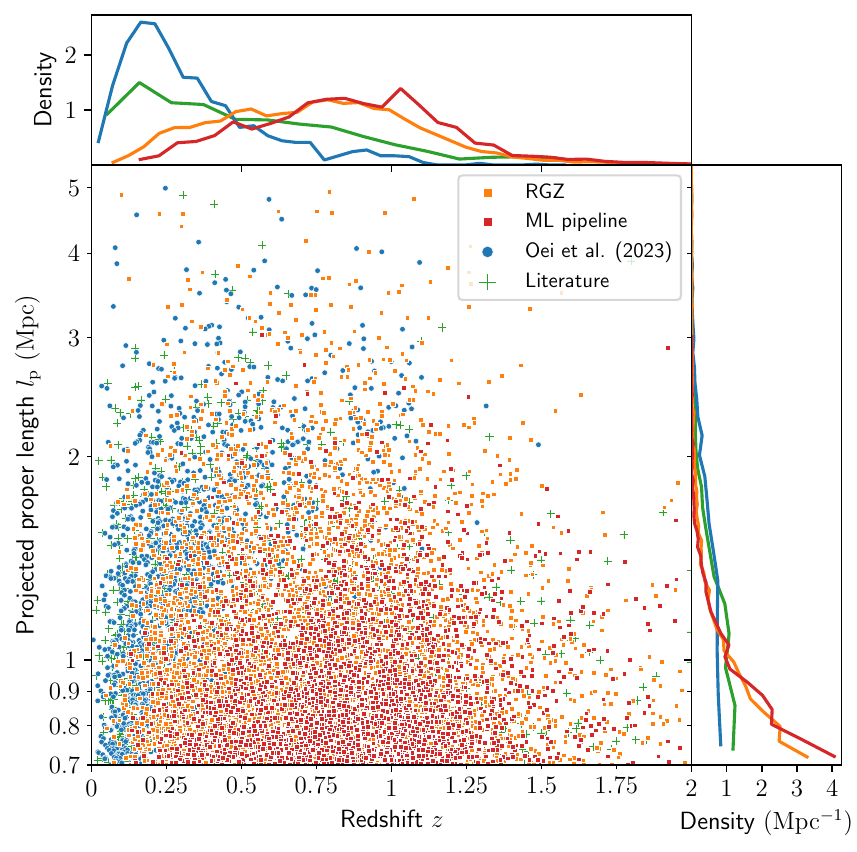}
\caption{Our sample of RGZ giants (orange squares) and ML pipeline giants (red squares) effectively complements the sample of giants with large angular lengths (blue dots) from the manual search of \citet{Oei2023Distribution}.
The remaining giants (green pluses) are from earlier literature, as specified in Sect.~\ref{sec:literature-sample}.
}
\label{fig:z_l}
\end{figure}

\begin{table*}[tb!]
      \caption[]{Statistics of the GRG samples that we discovered, confirmed, or aggregated.
      From left to right, the columns provide the number of giants in each sample, $N$, and the $10$th, the median,  and the $90$th percentile of the angular length $\phi$, redshift $z$, and projected proper length $l_{\mathrm{p}}$.}
         \label{tab:results}
    \centering
    \resizebox{\textwidth}{!}{%
    \begin{tabular}{l l l l l l l l l l l}
    \hline\hline
Sample & $N$ &  $\phi_\mathrm{10th}$ $(')$ & $\phi_\mathrm{median}$ $(')$ & $\phi_\mathrm{90th}$ $(')$ &  $z_\mathrm{10th}$ & $z_\mathrm{median}$ & $z_\mathrm{90th}$ &  $l_\mathrm{p,10th}$ $(\mathrm{Mpc})$ & $l_\mathrm{p,median}$ $(\mathrm{Mpc})$ & $l_\mathrm{p,90th}$ $(\mathrm{Mpc})$\\
\hline
ML pipeline & $2,722$ &  $1.50$ & $1.97$ & $3.09$&  $0.44$ & 0.87 & $1.28$ &  $0.72$ & $0.87$ & $1.29$\\
RGZ & $6,550$ &  $1.56$ & $2.19$ & $4.40$&  $0.31$ & 0.75 & $1.19$ &  $0.73$ & $0.91$ & $1.57$\\
Known giants & $11,585$ &  $1.57$ & $2.33$ & $5.70$&  $0.23$ & 0.72 & $1.19$ &  $0.73$ & $0.94$ & $1.68$\\
\hline\end{tabular}%
}
\end{table*}

As a result, the samples complement each other: the sample of \citet{Oei2023Distribution} is more complete at lower redshifts and higher projected lengths, while the RGZ and ML pipeline samples are more complete at higher redshifts and lower projected lengths.
Figure~\ref{fig:z_l} demonstrates this point, while Table \ref{tab:results} presents the corresponding statistics of the GRG samples.

For comparison of the $3\sigma$ lengths of the ML pipeline and RGZ giants to those in other surveys, we inform the reader that the central frequency and the average surface brightness threshold of the observations that we use are $\nu_\mathrm{obs} = 144\ \mathrm{MHz}$ and  $b_{\nu,\mathrm{th}} = 25\ \mathrm{Jy\ deg}^{-2}$ respectively.

\subsection{Estimating Bayesian model parameters}
\label{sec:methods_parameters}
After having refined our statistical GRG framework (Sect.~\ref{sec:theory}), and after having assembled the largest sample of giants yet (Sects.~\ref{sec:detection}--\ref{sec:literature-sample}), we combined both advances to perform inference of the length distribution, number density, and lobe volume-filling fraction of giants.

Given that our goal has been to infer properties of the full population of giants, rather than just of those currently observed, we included two main selection effects in our forward modelling.
As detailed in Sect.~\ref{sec:theorySurfaceBrightness}, we parametrised surface brightness selection with three parameters, which are free parameters of the model.
As detailed in Sect.~\ref{sec:theoryIdentification}, a second cause of selection is the imperfect operation of our three LoTSS DR2 search methods, all of which fail to identify a significant fraction of giants with lobe surface brightnesses \emph{above} the survey noise level.
We modelled this identification selection $p_{\mathrm{obs,ID}}$ with a set of logistic functions, regressed to GRG data.
We now provide details of this process.

\subsubsection{Identification probability functions}
\label{sec:empirical}
To estimate $p_\mathrm{obs,ID}(l_\mathrm{p},z)$ from data, we first selected all giants detected by the joint efforts of our machine learning pipeline, RGZ, and the manual, visual search of \citet{Oei2023Distribution}.
Next, we retained only those giants that are located in regions of the sky that have been scanned by all three searches.
This overlap region in principle corresponds to the full LoTSS DR2 coverage -- were it not for the fact that the search of \citet{Oei2023Distribution} skipped over the LoTSS DR1, which had already been scanned by \citet{Dabhade2020March}.
Therefore, the actual overlap region amounts to the LoTSS DR2 coverage with a spherical quadrangle removed, whose minimum and maximum right ascensions are $\alpha_\mathrm{min} = 160\degree$ and $\alpha_\mathrm{max} = 230\degree$ and whose minimum and maximum declinations are $\delta_\mathrm{min} = 45\degree$ and $\delta_\mathrm{max} = 56\degree$.
Appendix~\ref{app:sphericalQuadrangles} provides an explicit decomposition of our assumed LoTSS DR2 coverage -- and therefore implicitly of the overlap region -- in terms of disjoint spherical quadrangles.

\begin{figure*}
    \centering
    \begin{subfigure}{\columnwidth}
    \includegraphics[width=\columnwidth]{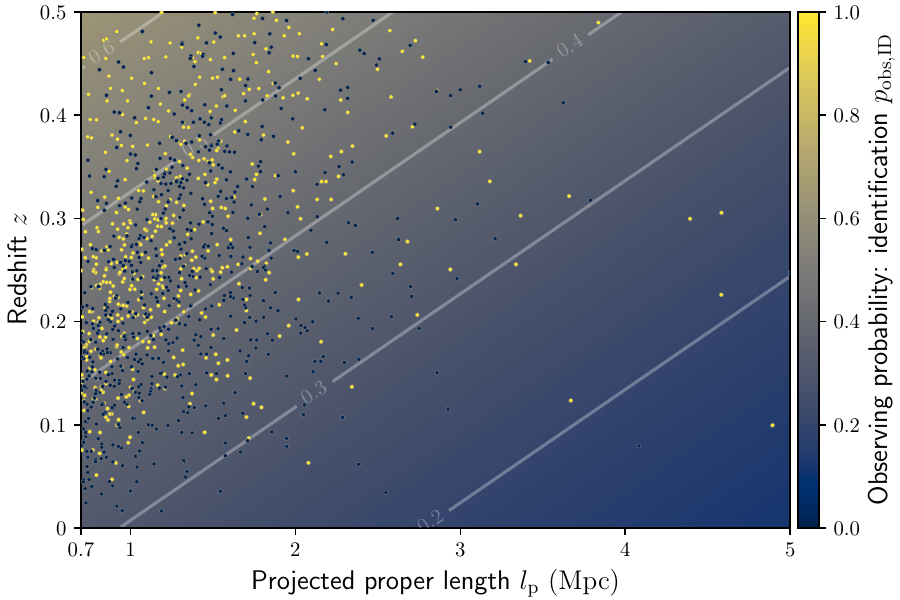}
    \end{subfigure}
    \begin{subfigure}{\columnwidth}
    \includegraphics[width=\columnwidth]{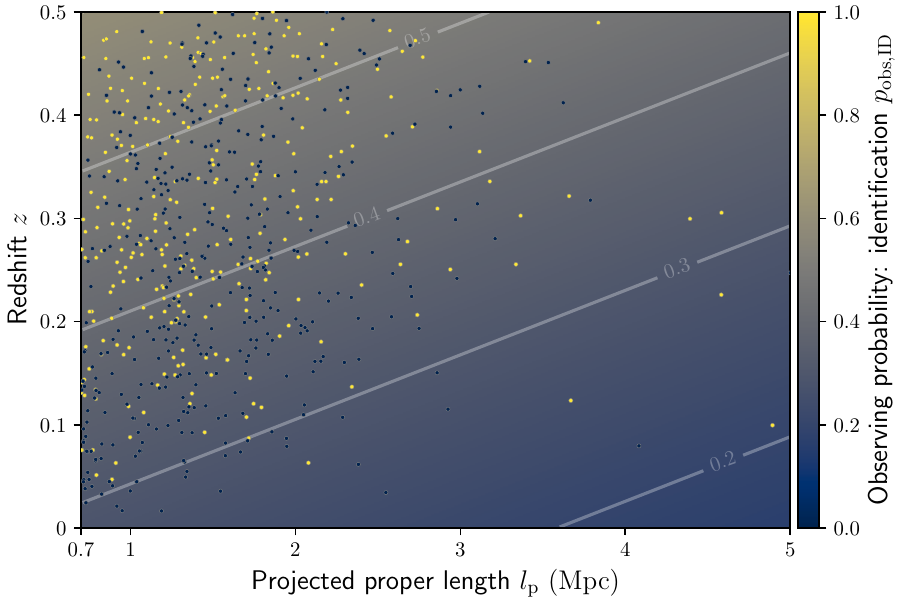}
    \end{subfigure}
    \begin{subfigure}{\columnwidth}
    \includegraphics[width=\columnwidth]{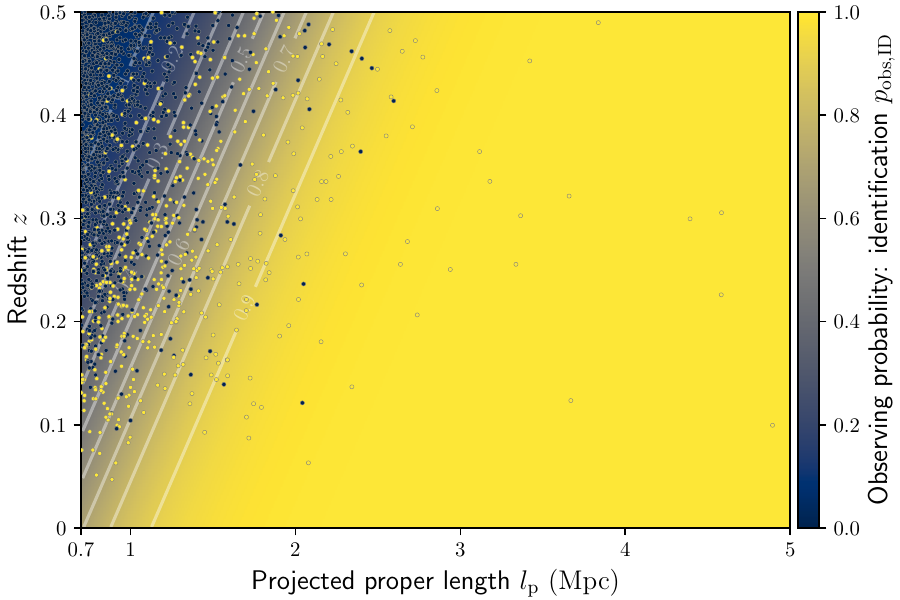}
    \end{subfigure}
    \begin{subfigure}{\columnwidth}
    \includegraphics[width=\columnwidth]{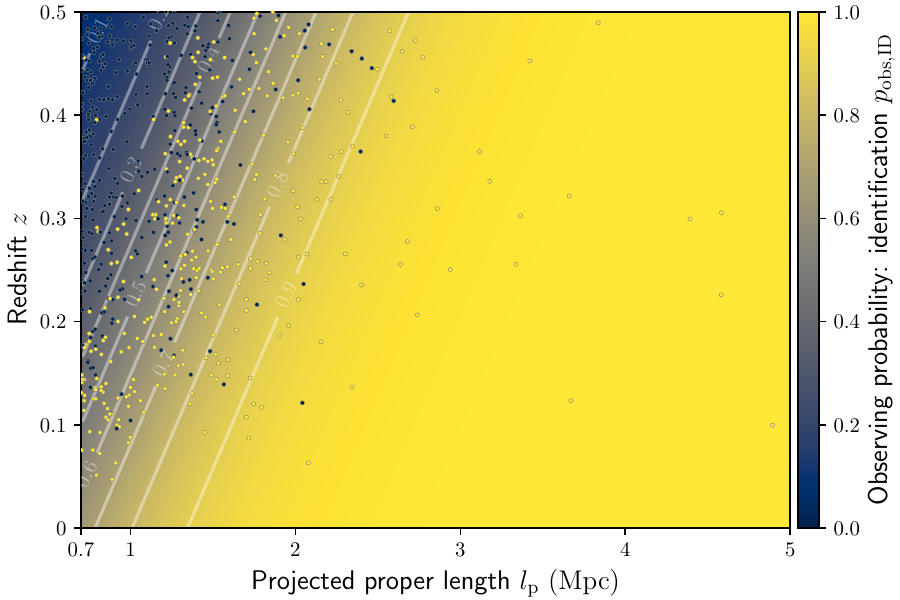}
    \end{subfigure}
    \begin{subfigure}{\columnwidth}
    \includegraphics[width=\columnwidth]{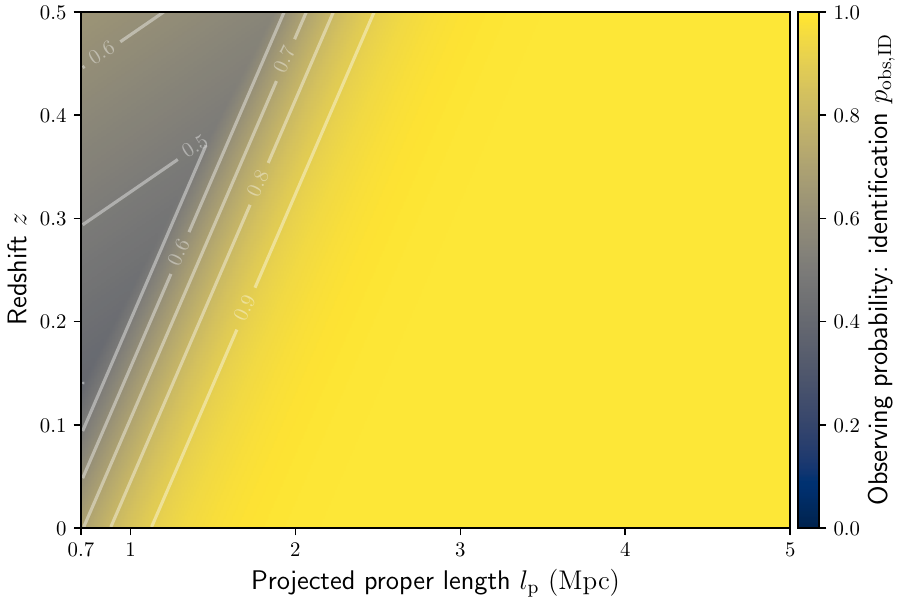}
    \end{subfigure}
    \begin{subfigure}{\columnwidth}
    \includegraphics[width=\columnwidth]{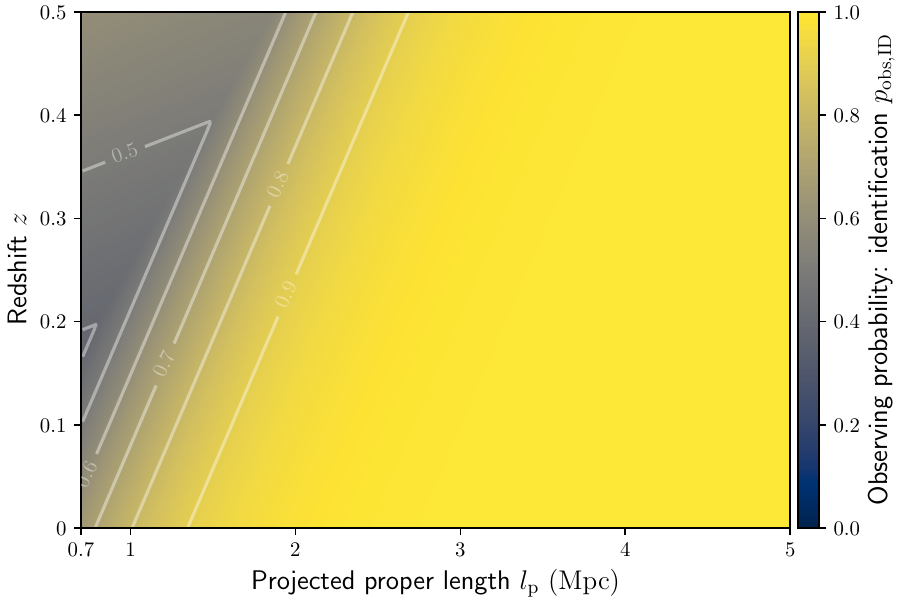}
    \end{subfigure}
    \caption{
    Overview of our determination of the probability to identify giants in the LoTSS DR2 with above-noise surface brightnesses, as a function of projected length and redshift -- through Radio Galaxy Zoo: LOFAR and our machine learning pipeline (top row), through the search of \citet{Oei2023Distribution} (middle row), and through these methods in unison (bottom row).
    Each of the upper four panels shows a binary logistic regression following the theory of Sect.~\ref{sec:theoryIdentification} and the practical considerations of Sect.~\ref{sec:empirical}.
    The left column shows results from all available data, whilst the right column shows results from rebalanced data.
    In our Bayesian inference, we used the latter results.}
    \label{fig:probabilityObservingID}
\end{figure*}

Some of the retained giants have been detected only in the combined RGZ--ML search, others have been detected only in the \citet{Oei2023Distribution} search, and yet others have been detected in both.
We note that, had it operated flawlessly, the combined RGZ--ML search would have detected all sources claimed by \citet{Oei2023Distribution} (or at least those that are genuine giants -- which should be the vast majority).
Therefore, by mapping the (in)ability of the RGZ--ML search to detect the giants of \citet{Oei2023Distribution} as a function of $l_\mathrm{p}$ and $z$, one can estimate the RGZ--ML search's identification probability function, $p_\mathrm{obs,ID,1}(l_\mathrm{p},z)$.
More precisely, for each giant detected by \citet{Oei2023Distribution}, we evaluated whether it was also detected in the RGZ--ML search, and stored a corresponding Boolean (that is to say, either $1$ or $0$).
We show these Booleans, at the $(l_\mathrm{p},z)$ coordinates of the giants they belong to, as yellow (representing $1$) and blue (representing $0$) dots in the top-left panel of Fig.~\ref{fig:probabilityObservingID}.
Viewing the Boolean at $(l_\mathrm{p},z)$ as a realisation of a Bernoulli RV with success probability $p = p_\mathrm{obs,ID,1}(l_\mathrm{p},z)$, we recognise the inference of the identification probability function as a binary logistic regression problem with two explanatory variables.
The background of Fig.~\ref{fig:probabilityObservingID}'s top-left panel shows the corresponding best fit.

By symmetry, this approach can be reversed to estimate the \citet{Oei2023Distribution} search's identification probability function, $p_\mathrm{obs,ID,2}(l_\mathrm{p},z)$.
Therefore, for each giant detected in the RGZ--ML search, we evaluated whether it was also detected by \citet{Oei2023Distribution}, and stored a corresponding Boolean.
In the same way as before, we show these Booleans in the middle-left panel of Fig.~\ref{fig:probabilityObservingID}.
The panel's background shows the best logistic fit.

We combine the two identification probability functions, $p_{\mathrm{obs,ID,1}}(l_\mathrm{p},z)$ and $p_{\mathrm{obs,ID,2}}(l_\mathrm{p},z)$, in point-wise fashion as to obtain a single function $p_{\mathrm{obs,ID}}(l_\mathrm{p},z)$.
To do so, we follow the minimal combination rule of Eq.~\ref{eq:identificationProbabilityCombinationMaximum}.

We remark that, by giving each Boolean in these logistic regressions an equal weight, the resulting functions are tuned to fit crowded regions of projected length--redshift parameter space best -- at the expense of accuracy in sparser regions.
To increase the accuracy of the functions for the parameter space at large, we performed a simple rebalancing step.
First, we calculated the mean number density in the parameter space given by $l_\mathrm{p} \in [0.7, 5\ \mathrm{Mpc}] \times [0, 0.5] \ni z$.
We then selectively subsampled the data in crowded regions, following the rule that the number density in each bin of width $0.5\ \mathrm{Mpc}$ and height $0.05$ should not exceed twice the mean number density of the entire parameter space.
We show the rebalanced data, alongside refitted logistic models, in the right column of Fig.~\ref{fig:probabilityObservingID}.
We report the rebalanced model coefficients in Table~\ref{tab:parametersConstants}, and treat them as constants during the Bayesian inference.

\subsubsection{Inference in practice}
\label{sec:inferencePractice}
In this work, we constrained the parameters of Sect.~\ref{sec:theory}'s GRG population model via a projected length--redshift histogram.
From our most extensive sample of giants, we selected those with $0.7\ \mathrm{Mpc} \eqqcolon l_\mathrm{p,GRG} < l_\mathrm{p} < 5.1\ \mathrm{Mpc}$ and $0 < z < z_\mathrm{max} \coloneqq 0.5$ that lie in the LoTSS DR2 coverage as specified in Appendix~\ref{app:sphericalQuadrangles}.
We did not include the giants from \citet{Oei2023Distribution} for which only a lower bound to the redshift is known.
This selection retained $2,685$ out of $11,585$ giants.
We used these giants to fill a histogram with bins of width $\Delta l_\mathrm{p} = 0.1\ \mathrm{Mpc}$ and $\Delta z = 0.02$.
We did not systematically explore the effect of these bin size parameters on the resulting inference.
However, the smaller one chooses the bins, the higher the numerical cost will be.
On the other hand, if the bins are chosen much larger than the typical scales over which the underlying observed projected length--redshift distribution\footnote{With the `underlying' observed projected length--redshift distribution, we mean the observed projected length--redshift distribution one would obtain in the limit of an infinite number of observed giants.} varies, then some ability to extract parameter constraints will be lost.

To compute the posterior distribution over the six parameters $\vec{\theta} = [\xi(l_\mathrm{p,GRG}), \Delta\xi, b_\mathrm{\nu,ref}, \sigma_\mathrm{ref}, \zeta, n_\mathrm{GRG}]$, we assumed a uniform prior and brute-force evaluated the likelihood function over a regular grid that covers a total of $2.1 \cdot 10^9$ parameter combinations.\footnote{This approach is feasible by virtue of the low numerical cost of each likelihood function evaluation.
Its main advantage is its simplicity: there are no parameters to tune that govern the method's convergence behaviour.
Once the model is expanded to include more parameters, or when selection effects with higher numerical cost are incorporated, more efficient (though more complicated) methods such as Markov chain Monte Carlo or nested sampling will become necessary.}
In doing so, we applied the Poissonian likelihood trick described in Appendix~\ref{app:likelihood}, which sped up our computations by one to two orders of magnitude.
Table~\ref{tab:parametersConstants} provides the parameter ranges for which we evaluated the likelihood (which coincide with their prior distribution ranges), alongside all of the model's constants and their assumed values.
Because each likelihood function evaluation can be computed independently of the others, the problem is fully parallelisable.
In practice, we distributed the ${\sim}10^4$ core-hours Python calculation over ${\sim}1500$ virtual cores, which were spread across ${\sim}20$ nodes of a computer cluster.
Next, we generated samples from the posterior distribution using rejection sampling \citep[e.g.][]{Rice2006}.
We subsequently used these samples to calculate probability distributions for derived quantities.\footnote{To calculate probability distributions over quantities that are a function of the parameters, such as the Local Universe GRG lobe VFF, $\mathcal{V}_\mathrm{GRG-CW}(z = 0)$, or the joint search completeness function $C$, we could in principle evaluate these quantities for each parameter combination of the aforementioned grid and weigh each grid point's result by the associated likelihood (or, equivalently, posterior probability).
However, some derived quantities are costly to compute, so that excessive evaluations should be avoided.}

\begin{table}[]
    \caption{
    Parameters and constants of GRG population forward model from Sect.~\ref{sec:theory} alongside their prior ranges and values, as used in the Bayesian inference presented in Sect.~\ref{sec:results}.
    The first six constants serve to define the quantitative meaning of the parameters and set the scope of the analysis.
    The other eleven constants are not arbitrary: they affect the likelihood function and posterior distribution for a given set of parameter definitions and scope.
    }
    \centering
    \begin{tabular}{l l l}
    \hline\hline
    Parameter & Uniform prior range & Explanation\\\hline
    $\xi(l_\mathrm{p,1} = l_\mathrm{p,GRG})$      & $[-3.5, -2]$ & Sect.~\ref{sec:theoryCurvedPowerLaw}\\
    $\Delta\xi$ & $[-3.5, -1.5]$ & Sect.~\ref{sec:theoryCurvedPowerLaw}\\
    $b_{\nu,\mathrm{ref}}$ & $[1, 100] \cdot \mathrm{Jy\ deg^{-2}}$ & Sect.~\ref{sec:theorySurfaceBrightness}\\
    $\sigma_\mathrm{ref}$ & $[0.5, 2]$ & Sect.~\ref{sec:theorySurfaceBrightness}\\
    $\zeta$ & $[-0.5, 0]$ & Sect.~\ref{sec:theorySurfaceBrightness}\\
    $n_\mathrm{GRG}$ & $[0, 50] \cdot (100\ \mathrm{Mpc})^{-3}$ & Sect.~\ref{sec:theoryNumberDensity}\\
    \hline
    & &\\
    \hline\hline
    Constant & Value & Explanation\\\hline
    $l_\mathrm{p,GRG}$      & $0.7\ \mathrm{Mpc}$ & Sect.~\ref{sec:intro}\\
    $l_\mathrm{p,1}$ & $0.7\ \mathrm{Mpc}$ & Sect.~\ref{sec:theoryCurvedPowerLaw}\\
    $l_\mathrm{p,2}$ & $5\ \mathrm{Mpc}$ & Sect.~\ref{sec:theoryCurvedPowerLaw}\\
    $l_\mathrm{ref}$ & $0.7\ \mathrm{Mpc}$ & Sect.~\ref{sec:theorySurfaceBrightness}\\
    $\nu_\mathrm{obs}$ & $144\ \mathrm{MHz}$ & Sect.~\ref{sec:theorySurfaceBrightness}\\
    $z_\mathrm{max}$ & $0.5$ & Sect.~\ref{sec:theorySelection}\\
    $\alpha$ & $-1$ & Sect.~\ref{sec:theorySurfaceBrightness}\\
    $b_{\nu,\mathrm{th}}$ & $25\ \mathrm{Jy\ deg^{-2}}$ & Sect.~\ref{sec:theorySurfaceBrightness}\\
    $\beta_{0,1}$ & $-1.0$ & Sect.~\ref{sec:theoryIdentification}\\
    $\beta_{0,2}$ & $-1.0$ & Sect.~\ref{sec:theoryIdentification}\\
    $\beta_{l_\mathrm{p},1}$ & $-0.1\ \mathrm{Mpc}^{-1}$ & Sect.~\ref{sec:theoryIdentification}\\
    $\beta_{l_\mathrm{p},2}$ & $2.4\ \mathrm{Mpc}^{-1}$ & Sect.~\ref{sec:theoryIdentification}\\
    $\beta_{z,1}$ & $2.8$ & Sect.~\ref{sec:theoryIdentification}\\
    $\beta_{z,2}$ & $-6.4$ & Sect.~\ref{sec:theoryIdentification}\\
    $\Delta l_\mathrm{p}$ & $0.1\ \mathrm{Mpc}$ & Sect.~\ref{sec:inferencePractice}\\
    $\Delta z$ & $0.02$ & Sect.~\ref{sec:inferencePractice}\\
    $\Omega$ & $1.62\ \mathrm{sr}$ & Appendix~\ref{app:sphericalQuadrangles}\\
    \hline
    \end{tabular}
    \label{tab:parametersConstants}
\end{table}

\section{Results}
\label{sec:results}
\begin{figure*}
    \centering
    \includegraphics[width=\textwidth]{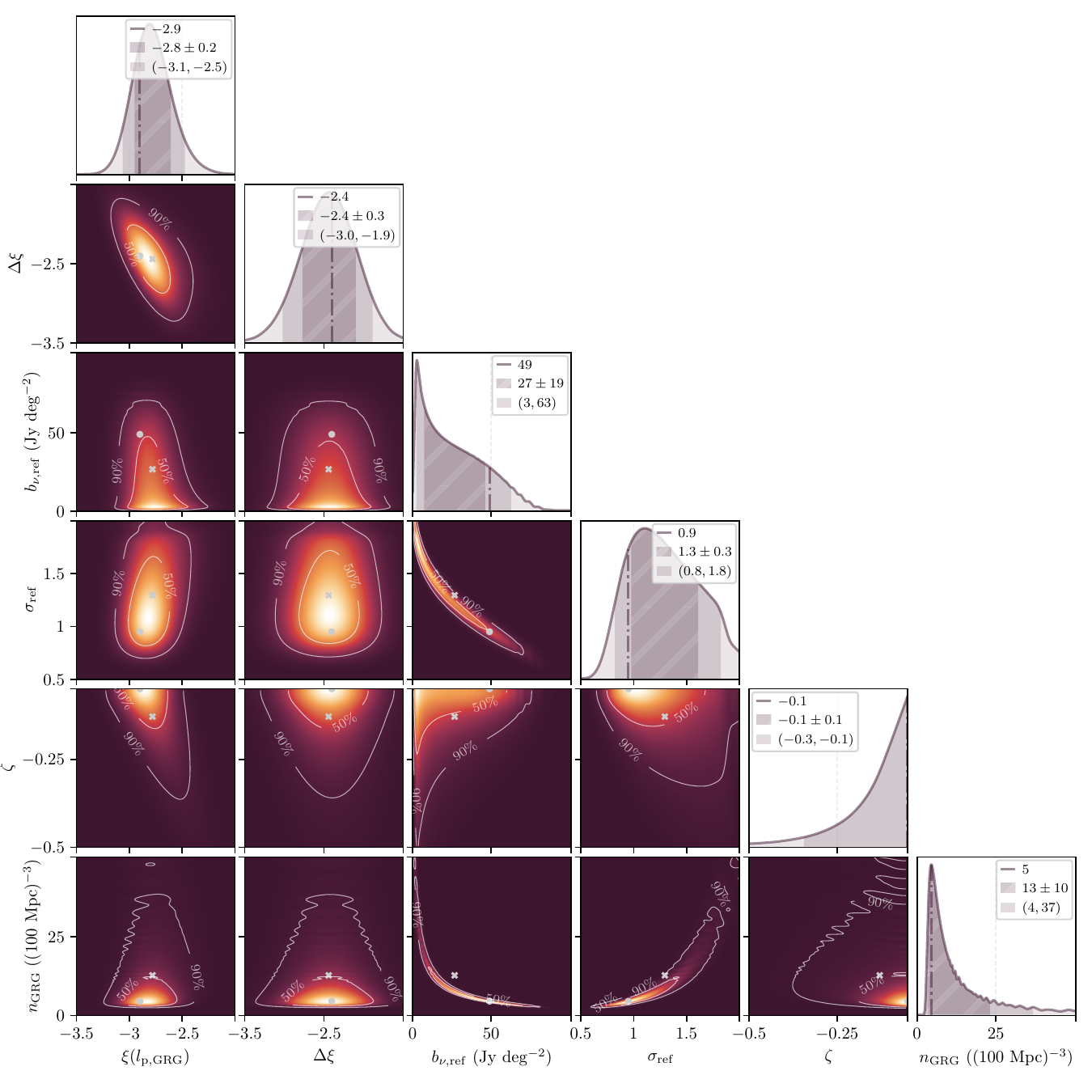}
    \caption{Likelihood function over $\vec{\theta} = [\xi(l_\mathrm{p,GRG}), \Delta\xi, b_{\nu,\mathrm{ref}}, \sigma_\mathrm{ref}, \zeta, n_\mathrm{GRG}]$, based on 2,685 projected lengths and redshifts of giants up to $z_\mathrm{max} = 0.5$.
    We show all two-parameter marginals of the likelihood function, with contours enclosing 50\% and 90\% of total probability.
    We mark the maximum likelihood estimate (MLE) values (grey dot) and the likelihood mean values (grey cross).
    The one-parameter marginals again show the MLE (dash-dotted line), a mean-centred interval of standard deviation--sized half-width (hashed region), and a median-centred 90\% credible interval (shaded region).}
    \label{fig:posterior}
\end{figure*}

By combining an unparalleled sample of giant radio galaxies with a rigorous forward model, we have produced a posterior distribution over parameters that characterise the intrinsic population of giants.
Figure~\ref{fig:posterior} summarises the posterior over parameter hexads $\vec{\theta} = [\xi(l_\mathrm{p,GRG}), \Delta\xi, b_{\nu,\mathrm{ref}}, \sigma_\mathrm{ref}, \zeta, n_\mathrm{GRG}]$ by means of its one- and two-dimensional marginal distributions.
In this section, we analyse our newfound parameter constraints.

\subsection{Length distribution of giant radio galaxies}
Radio galaxies enrich the IGM with magnetic fields, but giants -- given their megaparsec-scale reach -- appear uniquely capable of seeding the more remote regions of the Cosmic Web.
Consequently, scientific interest in quantifying the length distribution of giants has arisen from the possibility that giants contribute significantly to cosmic magnetogenesis.
The question at hand is deceivingly simple: how common are giants of various lengths?

As pointed out by \citet{Oei2023Distribution}, due to selection effects, the observed projected length distribution is not a reliable estimate of the \emph{true} projected length distribution.
Worse still, the relevant selection effects might not be quantitatively known a priori, requiring joint inference of the length distribution, and the selection effect parameters.
\citet{Oei2023Distribution} performed such a joint inference, and found that their data were consistent with an underlying population of giants with Pareto-distributed lengths, characterised by tail index $\xi = -3.4 \pm 0.5$.
In the current work, we have relaxed the assumption of perfect Paretianity, and explore whether the data are consistent with a curved power law PDF for the GRG projected proper length RV $L_\mathrm{p}\ \vert\ L_\mathrm{p} \geq l_\mathrm{p,GRG}$.
The marginals of Fig.~\ref{fig:posterior} suggest that they are -- in fact, the data strongly favour curvature, with a tail index at $l_\mathrm{p,1} \coloneqq l_\mathrm{p,GRG} \coloneqq 0.7\ \mathrm{Mpc}$ of $\xi(l_\mathrm{p,GRG}) = -2.8 \pm 0.2$ and a total increase in tail index up to $l_\mathrm{p,2} \coloneqq 5\ \mathrm{Mpc}$ of $\Delta\xi = -2.4 \pm 0.3$.
Given the small relative uncertainty on the latter value, our data appear inconsistent with perfect Paretianity ($\Delta \xi = 0$).
We note that our notion of `data' is different from that in \citet{Oei2023Distribution}: not only do we use more than a thousand additional giants, we also make more effective use of their redshift information.
For further discussion, see Sect.~\ref{sec:previousInference}.

It remains an open question whether giants can be understood as part of the ordinary radio galaxy population, or whether they evolve through qualitatively different physical processes.
As pointed out in Sect.~4.1.5 of \citet{Oei2023Distribution}, a curved power law PDF for $L_\mathrm{p}\ \vert\ L_\mathrm{p} \geq l_\mathrm{p,GRG}$ is consistent with a scenario in which giants share a broader length continuum with smaller radio galaxies.
More specifically, if the broader radio galaxy length distribution is approximately lognormal, as appears justifiable on statistical grounds, then $\xi$ should decrease throughout the distribution's right tail -- that is, throughout the GRG range.
Future research should determine whether such a unified non-giant RG--GRG scenario is also \emph{quantitatively} consistent with the decrease in $\xi$ we have inferred here.
In addition, our inferences of $\xi(l_\mathrm{p,GRG})$ and $\Delta\xi$ are important in constraining Sect.~\ref{sec:resultsVFF}'s GRG lobe volume-filling fraction.

\subsection{Number density of giant radio galaxies}
The extent to which giants have contributed to cosmic magnetogenesis depends on their intrinsic number density -- which need not necessarily be a constant, but could have evolved over time.
Observationally, giants are considered rare in comparison to smaller radio galaxies.
However, because giants are presumably strongly affected by surface brightness selection, this present-day observed rarity might not translate to an intrinsic rarity.
Excitingly, by forward modelling selection effects -- and in particular surface brightness selection -- we can constrain the intrinsic comoving GRG number density between $z = 0$ and $z = z_\mathrm{max}$, which we denote simply by $n_\mathrm{GRG}$.

The bottom-right one-dimensional marginal of Fig.~\ref{fig:posterior} shows a strongly skewed distribution for $n_\mathrm{GRG}$, with a marginal mean $\mathbb{E}[n_\mathrm{GRG}] = 13 \pm 10\ (100\ \mathrm{Mpc})^{-3}$ and a $95\%$ probability that $n_\mathrm{GRG} > 4\ (100\ \mathrm{Mpc})^{-3}$.
These number densities are a factor of order unity higher than those of \citet{Oei2023Distribution}, who inferred a marginal mean $\mathbb{E}[n_\mathrm{GRG}] = 4.6 \pm 2.4\ (100\ \mathrm{Mpc})^{-3}$ and a $90\%$ probability that $n_\mathrm{GRG} < 6.7\ (100\ \mathrm{Mpc})^{-3}$.

The joint marginal distribution of $n_\mathrm{GRG}$ and $b_{\nu,\mathrm{ref}}$ reveals a strong inverse relationship, whose origin is easy to grasp.
Models in which giants are relatively rare (i.e. with low $n_\mathrm{GRG}$) but with relatively mild surface brightness selection (i.e. with high $b_{\nu,\mathrm{ref}}$) are about as successful in reproducing the data-derived projected length--redshift histogram as models in which giants are relatively common (i.e. with high $n_\mathrm{GRG}$) but with relatively severe surface brightness selection (i.e. with low $b_{\nu,\mathrm{ref}}$).
The narrowness of the joint distribution also suggests that, if estimates of $b_{\nu,\mathrm{ref}}$ would reveal it to be $\gtrsim 10\ \mathrm{Jy\ deg^{-2}}$, it should be possible to break the degeneracy and accurately determine $n_\mathrm{GRG}$.

Recent work \citep{Oei2024GiantsWeb} suggests that the comoving number density of luminous, non-giant radio galaxies (LNGRGs), understood to have radio luminosities at $150\ \mathrm{MHz}$ of $l_\nu \geq 10^{24}\ \mathrm{W\ Hz^{-1}}$ and projected lengths $l_\mathrm{p} < l_\mathrm{p,GRG} \coloneqq 0.7\ \mathrm{Mpc}$, is $n_\mathrm{LNGRG} = 12 \pm 1\ (100\ \mathrm{Mpc})^{-3}$.
Our work suggests that giants might be comparably common.
If this is indeed the case, then the widespread belief that giants form a rare population of radio galaxies must be revised.

\subsection{Lobe volume-filling fraction of giant radio galaxies}
\label{sec:resultsVFF}
The present-day volume-filling fraction of the lobes of giants in clusters and filaments of the Cosmic Web, $\mathcal{V}_\mathrm{GRG-CW}(z=0)$, is not a parameter of our model, but rather a derived quantity.
As briefly discussed in Sect.~\ref{sec:inferencePractice}, we compute its probability distribution using the parameter hexads that we have obtained by rejection sampling from the posterior.
For each sampled hexad, we compute $\xi(l_\mathrm{p})$ using $\xi(l_\mathrm{p,GRG})$, $\Delta\xi$, and Eq.~\ref{eq:exponentFunction2}, then $f_{L_\mathrm{p}\ \vert\ L_\mathrm{p} \geq l_\mathrm{p,GRG}}(l_\mathrm{p})$ using Eq.~\ref{eq:GRGProjectedProperLengthCurved}, and finally $\mathcal{V}_\mathrm{GRG-CW}(z=0)$ using $n_\mathrm{GRG}$ and Eq.~\ref{eq:VFF}.

\begin{figure}
    \centering
    \includegraphics[width=\columnwidth]{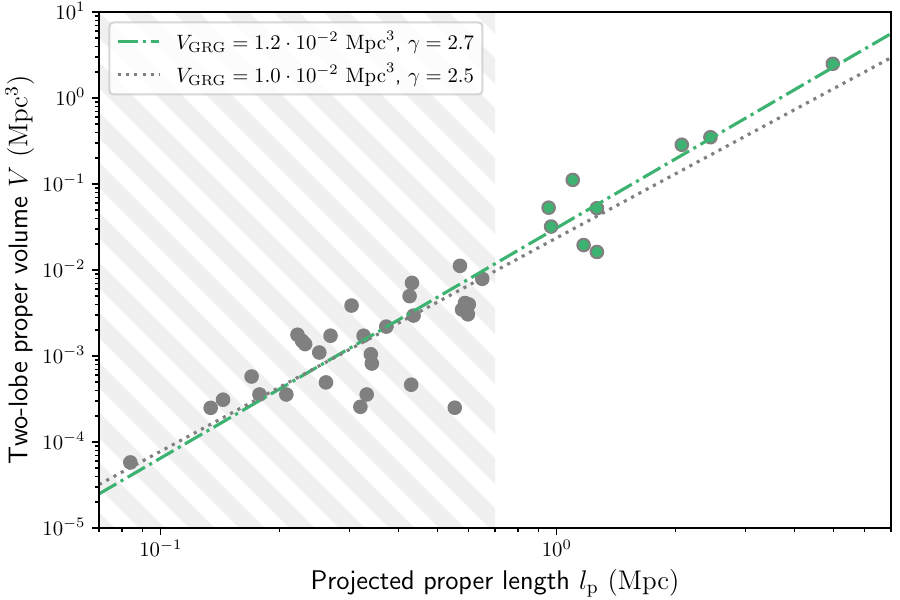}
    \caption{
    Empirical relation between projected proper length and two-lobe proper volume for RGs from \citet{Ineson2017}, \citet{Oei2022Alcyoneus}, and \citet{Oei2023NGC6185}.
    The power law--like trend motivates Eq.~\ref{eq:VFFModel}.
    The green fit is based on giants only, whilst the grey fit is based on all RGs.
    $V_\mathrm{GRG}$ is the mean two-lobe proper volume of the shortest possible giants (i.e. giants for which $l_\mathrm{p} = l_\mathrm{p,GRG}$), while $\gamma$ is the exponent of the power law. For self-similar RG growth, $\gamma = 3$.
    }
    \label{fig:projectedProperLengthTwoLobeVolume}
\end{figure}
To arrive at Eq.~\ref{eq:VFF}, we assumed in Eq.~\ref{eq:VFFModel} that RG projected proper lengths and two-lobe proper volumes obey a power law relation with scatter.
To investigate the validity of this assumption, we collected the projected proper lengths and two-lobe proper volumes of all Fanaroff--Riley II RGs in the `representative' sample of \citet{Ineson2017}.\footnote{We calculated projected proper lengths by summing the lobe tip distances of both lobes. For RGs for which \citet{Ineson2017}'s Table~5 reports data on only one lobe, we assumed that the other lobe has an identical lobe tip distance and volume.}
Among these RGs are just seven giants.
To more reliably probe the projected proper length--two-lobe proper volume relation for giants, we supplemented this sample with the giant of NGC 6185, the longest spiral galaxy--generated RG known \citep{Oei2023NGC6185}, and with Alcyoneus, the longest elliptical galaxy--generated RG known \citep{Oei2022Alcyoneus}.
\citet{Oei2023NGC6185} estimated that the giant of NGC 6185 measures $l_\mathrm{p} = 2.45 \pm 0.01\ \mathrm{Mpc}$ and has a two-lobe proper volume $V = 0.35 \pm 0.03\ \mathrm{Mpc}^3$.
Similarly, \citet{Oei2022Alcyoneus} estimated that Alcyoneus measures $l_\mathrm{p} = 4.99 \pm 0.04\ \mathrm{Mpc}$ and has a two-lobe proper volume $V = 2.5 \pm 0.3\ \mathrm{Mpc}^3$.
We show the empirical $l_\mathrm{p}$--$V$ relation for the resulting sample in Fig.~\ref{fig:projectedProperLengthTwoLobeVolume}.
Tentatively, we consider the assumed power law relation between projected proper length and two-lobe proper volume justified, although we warn that the sample size is small and that we have not made corrections for selection effects.
Through least-squares minimisation in log--log space, we obtained two best-fit power law relations: one for giants only (green line) and for all RGs (grey line).
In order to calculate Eq.~\ref{eq:VFF}, we adopted the parameters from the giant-based fit: $V_\mathrm{GRG} = 1.2 \cdot 10^{-2}\ \mathrm{Mpc}^3$ and $\gamma = 2.7$.

\begin{figure}
    \centering
    \includegraphics[width=\columnwidth]{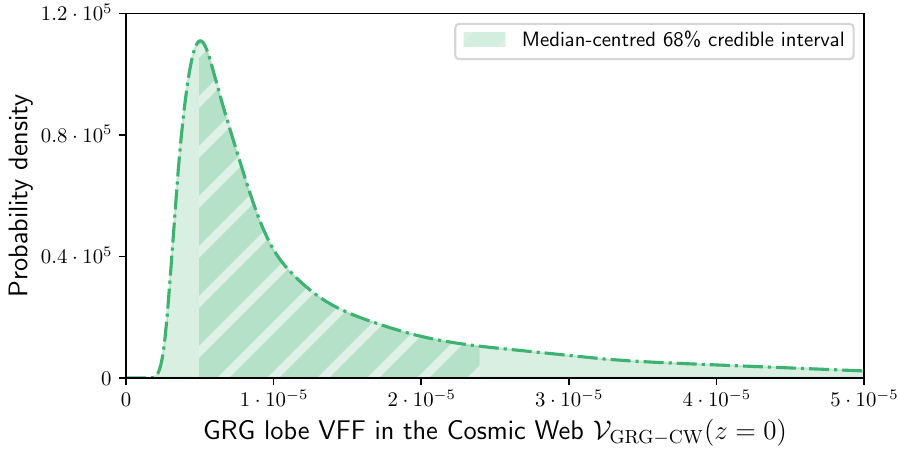}
    \caption{Posterior distribution for the instantaneous, present-day GRG lobe volume-filling fraction in clusters and filaments of the Cosmic Web, $\mathcal{V}_\mathrm{GRG-CW}(z = 0)$.}
    \label{fig:VFF}
\end{figure}

For each sampled hexad, we calculate $\mathbb{E}[L_\mathrm{p}^{\gamma}\ \vert\ L_\mathrm{p} \geq l_\mathrm{p,GRG}]$ using the law of the unconscious statistician.
The mean two-lobe volume of a giant is $\mathbb{E}[V\ \vert\ L_\mathrm{p} \geq l_\mathrm{p,GRG}](z=0) = 5.1 \pm 0.3\ \cdot 10^{-2}\ \mathrm{Mpc}^3$.
As in \citet{Oei2023Distribution}, we assume that clusters and filaments comprise $5\%$ of the Local Universe's volume \citep{ForeroRomero2009}: $\mathcal{V}_\mathrm{CW}(z=0) = 5\%$.
We show the resulting posterior distribution over $\mathcal{V}_\mathrm{GRG-CW}(z = 0)$ in Fig.~\ref{fig:VFF}.
This probability distribution inherits its skewness from the skewed marginal of $n_\mathrm{GRG}$.
We find $\mathcal{V}_\mathrm{GRG-CW}(z = 0) = 9\substack{+15\\-4} \cdot 10^{-6}$ and a posterior mean and standard deviation of $\mathcal{V}_\mathrm{GRG-CW}(z = 0) = 1.4 \pm 1.1 \cdot 10^{-5}$.
These results appear statistically consistent with that of \citet{Oei2023Distribution}: $\mathcal{V}_\mathrm{GRG-CW}(z=0) = 5\substack{+8\\-2}\cdot 10^{-6}$.

While they appear low at first sight, we speculate that these numbers are consistent with a scenario in which giants contribute significantly to cosmic magnetogenesis.
To see why, we first note that the number of giants that have ever existed might exceed those that exist now by two orders of magnitude: giants are actively powered for ${\sim}10^1$--$10^3\ \mathrm{Myr}$ \citep[e.g.][]{Hardcastle2018, Gurkan2022, Oei2022Alcyoneus, Dabhade2023}, the Universe is ${\sim}10^4\ \mathrm{Myr}$ old, and one may assume -- in the absence of clear evidence to the contrary -- that the comoving number density of giants $n_\mathrm{GRG}$ has remained roughly constant over cosmic time.\footnote{The dynamics of RGs are different at early epochs; reasons include a less developed large-scale structure but higher mean cosmic densities, higher AGN cold gas accretion and galaxy merger rates \citep[e.g.][]{OLeary2021}, and more severe inverse Compton energy losses to the CMB \citep[e.g.][]{Hardcastle2018}. Giants possibly representative of early epochs are discussed in e.g. \citet{Mahato2022}.}
Second, a likely large fraction of all giants that existed throughout cosmic history lived when the Universe was significantly smaller.
More specifically, all giants at $z \geq z_{10} \coloneqq \sqrt[3]{10}-1 \approx 1.15$ lived when the Universe's volume was at least an order of magnitude smaller; this period covers ${\sim}38\%$ of all cosmic time.
If the mean two-lobe proper volume of giants $\mathbb{E}[V\ \vert\ L_\mathrm{p} 
\geq l_\mathrm{p,GRG}]$ and the VFF of clusters and filaments $\mathcal{V}_\mathrm{CW}$ have remained roughly constant over the redshift range $[0, z_{10}]$, then by Eq.~\ref{eq:VFF} we have $\mathcal{V}_\mathrm{GRG-CW}(z = z_{10}) = 10 \cdot \mathcal{V}_\mathrm{GRG-CW}(z = 0)$.\footnote{Based on Millennium simulations, which are dark matter--only, \citet{Cautun12014} suggest that $\mathcal{V}_\mathrm{CW}$ may instead have decreased from ${\sim}10\%$ at $z = z_{10}$ to ${\sim}5\%$ at $z = 0$. This would imply that $\mathcal{V}_\mathrm{GRG-CW}(z = z_{10})$ is a factor ${\sim}2$ smaller than we claim here.}
Third, buoyant lobes might deposit magnetic fields in their wake, while diffusion, turbulence, and merger and accretion shocks from large-scale structure formation might have spread the contents of GRG lobes further through the IGM \citep[e.g.][]{Ensslin2003}.
The typical extent of a GRG lobe along a single spatial dimension, $\ell$, can be considered to be
\begin{align}
\ell(z) \coloneqq \sqrt[3]{\frac{\mathbb{E}[V\ \vert\ L_\mathrm{p} \geq l_\mathrm{p,GRG}](z)}{2}},
\end{align}
meaning that $\ell(z = 0) = 0.295 \pm 0.006\ \mathrm{Mpc} \sim 10^{-1}\ \mathrm{Mpc}$.
If, after jet fuelling stops, lobes rise buoyantly to the edges of filaments, their total traversed path length will be ${\sim}10^0\ \mathrm{Mpc}$ \citep[e.g.][]{Gheller2019}.
The columns through which GRG lobes have risen might therefore have a volume that is an order of magnitude larger than the lobes' own.
Taken together, these three effects could render the present-day VFF of magnetic fields that were once contained in the lobes of giants higher than $\mathcal{V}_\mathrm{GRG-CW}(z=0)$ by four (i.e. $2 + 1 + 1$) or more orders of magnitude.
This, in turn, suggests a significant astrophysical seeding potential.
For instance, assuming four orders of magnitude, ${\sim}10\%$ of the volume of today's Cosmic Web should have been magnetised by giants.

We finally point out that giant-induced IGM magnetic fields could have strengths consistent with observational constraints.
At the moment, the lowest magnetic field strengths measured in giant radio galaxy lobes, as inferred from images of Alcyoneus and the giant generated by NGC 6185 assuming the equipartition or minimum energy condition, are $400$--$500\ \mathrm{nG}$ \citep{Oei2022Alcyoneus, Oei2023NGC6185}.
If such field strengths would be typical, and buoyancy and diffusion lowers the density of field lines by an order of magnitude, then the typical giant-induced IGM field strength would be ${\sim}10\ \mathrm{nG}$.
This is in agreement with recent radio estimates and limits \citep[e.g. Table 1 of][]{Vazza2021}.
We note that this argument ignores possibly significant amplification and decay mechanisms, such as turbulent amplification and decay.

\section{Discussion}
\label{sec:discussion}
Below, we discuss how our ML pipeline and GRG population inference compare to earlier work.

\subsection{Comparison with previous machine learning search techniques}
\label{sec:past}
\citet{Proctor2016} applied an ML approach to search for GRG candidates by looking for likely pairs of (unresolved) radio lobes with the required angular length in the NRAO VLA Sky Survey \citep[NVSS;][]{Condon1998}. 
For this radio source component association problem, \citet{Proctor2016} trained an oblique classifier \citep[a type of decision tree ensemble;][]{Murthy1993}, using six source finder--derived features on $51,195$ pairs of radio components, $48$ of which were verified giants.
This method proved to be useful under the assumption that giants generally appear as an isolated pair of unresolved radio blobs, which is the case for NVSS with its $45''$ resolution and $450$ $\mathrm{\mu Jy}$ $\mathrm{beam^{-1}}$ sensitivity.
\citet{Dabhade2020October} visually inspected the $1,600$ GRG candidates presented by \citet{Proctor2016} and confirmed $151$ giants, which implies a $9\%$ precision for the GRG candidate predictions.
However, \citet{Proctor2016} expect that giants with resolved lobes -- which rule-based source finders often incorrectly break down into multiple separate sources -- require a different approach, and virtually all GRG lobes in LoTSS are resolved.\footnote{
As all giants have angular lengths $\phi \geq 1.3'$, they cover at least $13$ LoTSS $6''$ beams. This suggests that a single GRG lobe will cover multiple beams, too.}
It works in our favour that the convolutional neural network in our ML pipeline (Sect.~\ref{sec:association}) was specifically designed to use the morphology of the resolved, extended emission as a cue for the radio source component association.
Furthermore, as the source suggestions from our ML pipeline include optical host identifications, the candidates that we inspected not only have the required angular length but also have a host galaxy and corresponding redshift estimate assigned.
This allows us to visually inspect only those radio sources that fulfil the projected proper length $l_\mathrm{p,GRG} := 0.7$ Mpc requirement. %
Overall, our ML pipeline has a precision of $47\%$ for the GRG candidates that it suggests.

\subsection{Comparison with previous inference strategies}
\label{sec:previousInference}
Compared to the approach of \citet{Oei2023Distribution}, our approach makes better use of the redshift information available for each giant.
More specifically, we use the redshifts to make a `redshift-resolved' observed projected length histogram, while \citet{Oei2023Distribution} only compared a `redshift-collapsed' distribution of observed projected lengths to forward model predictions of $L_\mathrm{p,obs}\ \vert\ L_\mathrm{p,obs} \geq l_\mathrm{p,GRG}$.
Effectively, \citet{Oei2023Distribution} therefore used for each giant only \emph{Boolean} redshift information, $\mathbb{I}(z_i < z_\mathrm{max})$: that is, a truth value indicating whether or not the giant with index $i$ resides at a redshift below the maximum considered value.

In addition, our work changed the comoving number density of giants, $n_\mathrm{GRG}$, from a derived quantity to a model parameter.
This approach acknowledges the fact that the observed number of giants, either for a specific projected length--redshift bin or for the parameter space in its entirety, scales linearly with $n_\mathrm{GRG}$ (if the selection effects remain the same).
Therefore, there is intrinsic population information contained in the observed \emph{number} of giants.
However, by comparing predicted and observed \emph{probability distributions} only, \citet{Oei2023Distribution} did not exploit this fact.

\subsection{Future work}
\label{sec:future}
With the advent of large-scale, sensitive, low-frequency sky surveys such as the LoTSS, the Evolutionary Map of the Universe survey \citep[EMU;][]{Norris2011}, and the arrival of next-generation instruments such as the SKA \citep{Dewdney2009} and the DSA-2000 \citep[e.g.][]{Hallinan2019, Connor2022} later this decade, opportunities will arise to detect many more giants than have been found hitherto.
It is therefore likely that automated approaches to giant finding and host association will become only more relevant in the future.

Regarding our own machine learning--based pipeline, there is significant room to improve both the radio component association and the host association. Visual inspection indicated a precision of $47\%$ and the empirically determined $p_{\mathrm{obs,ID}}$ in Fig. \ref{fig:probabilityObservingID} showed that even in combination with the RGZ sample, the ML pipeline recall does not surpass $70\%$.
Sensible paths to improve the radio component association within the ML pipeline architecture include switching from rectangular bounding box--based object detection (the Fast R-CNN used in this article) to pixel-based instance segmentation and using a larger convolutional backbone \citep[e.g.][]{Liu2022convnet, Wright2010} or a transformer-based backbone \citep[e.g.][]{Liu2021, Zhang2022, Li2022}.
\citet{Mostert2022} conclude that a larger convolutional neural network is not effective unless one also significantly increases the quantity of high-quality training data, and in general, transformers require even more training data than convolutional neural networks \citep[e.g.][]{Wang2022}.
To that extent, adding a filtered version\footnote{For example, by identifying a handful of very active and expert volunteers and increasing the weight of their votes.} of the available LoTSS DR2 RGZ annotations \citep{Hardcastle2023} to the training data can be considered.
Furthermore, assembling a joined data set encompassing the (labelled) survey data of other low frequency radio telescopes can be considered.
Pretraining on this data set can benefit radio galaxy component association, host identification and morphological classification tasks across the board.

Finally, there appear to be clear opportunities to make the population-based forward model presented in Sect.~\ref{sec:theory} more accurate.
For example, at present, we have neglected photometric redshift uncertainties; however, the consequences of these uncertainties appear perfectly possible to forward model.
One such currently ignored consequence is Eddington bias: as RGs with projected lengths $l_\mathrm{p} = 0.6\ \mathrm{Mpc}$ are intrinsically more common than those with projected lengths $l_\mathrm{p} = 0.8\ \mathrm{Mpc}$, redshift error--induced projected length errors have the net effect of falsely raising the number of supposed giants with projected lengths near $l_\mathrm{p,GRG} \coloneqq 0.7\ \mathrm{Mpc}$.
This effect could contaminate the inference of $\xi(l_\mathrm{p,GRG})$.
Somewhat more challenging, but plausibly of greater value, would be a further exploration of how surface brightness selection is effectively modelled.
A major focus of such an exploration would be to analyse the surface brightness characteristics of hitherto discovered giants.
As the masked cutouts of Fig.~\ref{fig:reassess} suggest, the machine learning--based pipeline described in this work offers the exciting potential to amass -- fully automatically -- surface brightness properties for thousands of giants.
The availability of such properties for a large fraction of observed giants also allows one to fit the forward model to an observed projected length--redshift--surface brightness histogram, rather than to an observed projected length--redshift histogram only.
It is highly likely that adding another dimension to the data yields tighter parameter constraints.
To make the identification probability functions of Fig.~\ref{fig:probabilityObservingID} more accurate, it appears promising to have an expert visually (and exhaustively, i.e. without imposing angular length thresholds) comb through a small representative region of LoTSS DR2 in search of giants.
The resulting data set would provide a better basis for determining the identification probability functions than the RGZ--ML or \citet{Oei2023Distribution} data sets used in this work.
We note that the Bo\"otes LOFAR Deep Field search of \citet{Simonte2022} does not appear suited for this purpose, as the increased depth of this field renders it unrepresentative of LoTSS DR2 as a whole.
Finally, the model could be expanded in an attempt to measure cosmological evolution of, for example, $n_\mathrm{GRG}$.
However, we note that adding additional parameters to the model necessitates adopting more efficient inference techniques, such as Markov chain Monte Carlo or nested sampling.
The associated numerical gain would, in part, be negated by losing the speed-up associated to the likelihood trick of Appendix~\ref{app:likelihood}.

To determine the instantaneous VFF of GRG lobes in the Cosmic Web, $\mathcal{V}_\mathrm{GRG-CW}(z)$, one needs to calculate the mean GRG two-lobe proper volume $\mathbb{E}[V\ \vert\ L_\mathrm{p} \geq l_{\mathrm{p,GRG}}](z)$.
To estimate the latter quantity, we have proposed to leverage the apparent power law relation between projected length and two-lobe volume shown in Fig.~\ref{fig:projectedProperLengthTwoLobeVolume}.
However, the fit in this work is based on data of just nine giants.
To improve this situation, we recommend expanding the capabilities and automating the parametric Bayesian lobe volume estimation method introduced by \citet{Oei2022Alcyoneus, Oei2023NGC6185}.
This method could then be applied to thousands of our ML pipeline's masked cutouts, such as the one in Fig.~\ref{fig:reassess}.
This effort would increase the number of giants on which our fit of the projected length--two-lobe volume power law is based by two to three orders of magnitude.

\section{Conclusions}
\label{sec:conclusions}
In this work, we concatenated an existing crowd-sourced radio--optical catalogue, a new ML pipeline to automate radio--optical catalogue creation, and a Bayesian forward model to build a next-generation giant radio galaxy discovery and characterisation machine.
Applying this setup to the LOFAR Two-metre Sky Survey, we uncovered thousands of previously unknown giants, confirmed thousands of GRG candidates, and constrained the properties of the underlying population.
\begin{enumerate}
    \item The LoTSS is an ongoing sensitive, high-resolution, low-frequency radio survey whose second data release (DR2) covers $27\%$ of the northern sky.
    As the number of detected sources already ranges in the millions, it has become unfeasible (at least for small scientific teams) to conduct manual visual searches for giants, in particular for those with angular lengths close to the lower limit of $1.3'$.
    \item To address this challenge, we scanned all $841$ LoTSS DR2 pointings -- which together cover more than 5,000 square degrees of the northern sky -- with an ML pipeline that crucially includes the convolutional neural network of \citet{Mostert2022}, designed for the association of radio components for highly resolved radio galaxies, and an adapted version of the automated optical host galaxy identification heuristic developed by \citet{Barkus2022}.
    Used as a GRG detection system, our ML pipeline has a precision of $47\%$, a significant improvement over the $9\%$ precision obtained using the previously best ML GRG detection model \citep{Proctor2016, Dabhade2020October}.
    We merged the resulting giant candidate sample with that of the RGZ citizen science campaign \citep{Hardcastle2023}, homogenised the angular lengths, and subjected the candidates to a visual quality check.
    The result is a sample of more than 8,000 newly confirmed giants, of which a large fraction are considered genuine beyond reasonable doubt.
    More than $10^4$ unique giants have now been identified and published.
    \item We expand the population-based statistical forward model of \citet{Oei2023Distribution} designed to constrain the geometric properties of giants.
    In particular, by modelling the PDF of the radio galaxy projected length RV $L_\mathrm{p}$ as a curved power law, we automatically also model the PDF of the \emph{giant} radio galaxy projected length RV $L_\mathrm{p}\ \vert\ L_\mathrm{p} \geq l_\mathrm{p,GRG}$ as a curved power law.
    We assume that these projected length distributions do not undergo intrinsic evolution between cosmological redshifts of $z = z_\mathrm{max}$ and $z = 0$, and likewise assume an intrinsically constant comoving GRG number density throughout this redshift range.
    We model surface brightness selection by assuming a lognormal lobe surface brightness distribution at the survey's central frequency $\nu_\mathrm{obs}$, which is valid for radio galaxies of intrinsic proper length $l_\mathrm{ref}$ at redshift $z = 0$.
    We relate lobe surface brightness distributions for radio galaxies of other lengths and at other redshifts to this reference distribution.
    In addition, we model selection caused by the imperfect ability of search methods to identify all in-principle identifiable giants.
    For this purpose, we use logistic functions of projected length $l_\mathrm{p}$ and redshift $z$.
    \item We then sought to identify all model parameter hexads that can reproduce the projected length--redshift histogram of the joint RGZ--ML--\citet{Oei2023Distribution} LoTSS DR2 GRG sample.
    Through a simple Poissonian likelihood and a uniform prior distribution, we constructed a posterior distribution over the model parameters.
    By confronting the model with an observed projected length--redshift histogram, rather than with an observed projected length distribution only (as has been done in \citet{Oei2023Distribution}), we obtain tighter parameter constraints.
    \item We find evidence in support of the claim that the projected lengths of giant radio galaxies follow a curved power-law PDF, whose tail index is equal to $\xi(l_\mathrm{p,GRG}) = -2.8 \pm 0.2$ at $l_{\mathrm{p},1} = l_\mathrm{p,GRG} \coloneqq 0.7\ \mathrm{Mpc}$ and increases by $\Delta\xi = -2.4 \pm 0.3$ (i.e. decreases by $2.4 \mp 0.3$) in the projected length interval leading up to $l_{\mathrm{p},2} = 5\ \mathrm{Mpc}$.
    The predicted median lobe surface brightness at $\nu_\mathrm{obs} = 150\ \mathrm{MHz}$, $l_\mathrm{ref} = 0.7\ \mathrm{Mpc}$, and $z = 0$ is equal to $b_{\nu,\mathrm{ref}} = 30 \pm 20\ \mathrm{Jy\ deg^{-2}}$.
    This surface brightness level is lower than previously thought.
    Tight degeneracies resembling inverse relations exist between $b_{\nu,\mathrm{ref}}$ and the reference surface brightness dispersion measure $\sigma_\mathrm{ref}$, and between $b_{\nu,\mathrm{ref}}$ and the GRG number density $n_\mathrm{GRG}$.
    The latter relation suggests that giants might be more common than previously thought.
    At $n_\mathrm{GRG} = 13 \pm 10\ (100\ \mathrm{Mpc})^{-3}$, giants appear to be of an abundance comparable to that of luminous \emph{non}-giant radio galaxies.
    We conclude that, at any moment in time, a significant fraction of the radio galaxy population is in a GRG phase.
    As an immediate consequence, the fraction of radio galaxies that end their lives as giants must be even higher.
    \item Finally, we generate a posterior distribution for the instantaneous volume-filling fraction of GRG lobes in clusters and filaments of the Cosmic Web, $\mathcal{V}_\mathrm{GRG-CW}(z=0)$ -- a key statistic required for determining the cosmic magnetogenesis potential of giants.
    We find $\mathcal{V}_\mathrm{GRG-CW}(z=0) = 1.4 \pm 1.1 \cdot 10^{-5}$.
    The mean two-lobe proper volume of a giant is $\mathbb{E}[V\ \vert\ L_\mathrm{p} \geq l_{\mathrm{p,GRG}}](z=0) = 5.1 \pm 0.3 \cdot 10^{-2}\ \mathrm{Mpc}^3$.
    If a GRG population similar to that in the Local Universe has existed for most of the lifetime of the Universe, and IGM mixing processes are significant, then it appears possible that magnetic fields originating from giants permeate throughout significant (${\gtrsim}10\%$) fractions of today's Cosmic Web.
\end{enumerate}
Using modern automation and inference techniques (which still leave significant room for future improvements), we conducted the most detailed study of the abundance and geometrical properties of giants to date.
These cosmic colossi may provide a previously underappreciated contribution to astrophysical magnetogenesis.

\begin{acknowledgements}
The full GRG catalogue with host identifications %
will soon be available on the VizieR catalogue service hosted by the Centre de Données astronomiques de Strasbourg (CDS).
M.S.S.L. Oei and R.J. van Weeren acknowledge support from the VIDI research programme with project number 639.042.729, which is financed by the Dutch Research Council (NWO).
M.S.S.L. Oei also acknowledges support from the CAS--NWO programme for radio astronomy with project number 629.001.024, which is financed by the NWO.
Finally, M.S.S.L. Oei acknowledges support from the ERC Starting Grant ClusterWeb 804208.
L. Alegre is grateful for support from the UK STFC via CDT studentship grant ST/P006809/1.
M.J. Hardcastle acknowledges support from the UK STFC [ST/V000624/1].
B. Barkus is grateful for support from the UK STFC.
We would like to thank Huib Intema for enabling the cross-institute collaboration on the Leiden Observatory computer infrastructure.
We would like to thank Frits Sweijen for coding the very useful \url{https://github.com/tikk3r/legacystamps}.
This research has made use of the Python \texttt{astropy} package \citep{astropy}; the VizieR catalogue access tool \citep{ochsenbein2000vizier}, CDS, Strasbourg, France (DOI: 10.26093/cds/vizier); and the `Aladin Sky Atlas' developed at CDS, Strasbourg Observatory, France \citep{Bonnarel2000, Boch2014}.
LOFAR data products were provided by the LOFAR Surveys Key Science project (LSKSP; \url{https://lofar-surveys.org/}) and were derived from observations with the International LOFAR Telescope (ILT). LOFAR (van Haarlem et al. 2013) is the Low Frequency Array designed and constructed by ASTRON. It has observing, data processing, and data storage facilities in several countries, which are owned by various parties (each with their own funding sources), and which are collectively operated by the ILT foundation under a joint scientific policy. The efforts of the LSKSP have benefited from funding from the European Research Council, NOVA, NWO, CNRS-INSU, the SURF Co-operative, the UK Science and Technology Funding Council and the Jülich Supercomputing Centre.
This publication uses data generated via the Zooniverse.org platform, development of which is funded by generous support, including a Global Impact Award from Google, and by a grant from the Alfred P. Sloan Foundation.
\end{acknowledgements}

\bibliographystyle{aa} %
\bibliography{find_giants} %
\begin{appendix}
\section{Curved power law PDF for $L$}
\label{app:theory}
In Sect.~\ref{sec:theoryCurvedPowerLaw}, we have started modelling the geometry of radio galaxies at the level of the projected proper length RV $L_\mathrm{p}$.
While algebraically easier -- when curved power laws are considered, at least -- this approach is more limited than starting the forward model at the level of the intrinsic proper length RV $L$.
In this appendix, we calculate the distribution of $L_\mathrm{p}$ upon modelling $L$ with a curved power law.
Let us assume that, for $l \geq l_\mathrm{min}$,
\begin{align}
    f_L(l) \propto \left(\frac{l}{l_\mathrm{min}}\right)^{\xi(l)},
\end{align}
where $\xi(l) = al+b$.
We now use the identity that for $f(x) = \left(\frac{x}{c}\right)^{ax+b}$, one finds
\begin{align}
    \frac{\mathrm{d}f(x)}{\mathrm{d}x} = \left(\frac{x}{c}\right)^{ax+b}\left(a \ln{\frac{x}{c}} + a + \frac{b}{x}\right) = f(x)\left(a \ln{\frac{x}{c}} + a + \frac{b}{x}\right).
\end{align}
Therefore,
\begin{align}
    \frac{\mathrm{d}f_L(l)}{\mathrm{d}l} = f_L(l)\left(a \ln{\frac{l}{l_\mathrm{min}}} + a + \frac{b}{l}\right),
\end{align}
and
\begin{align}
    \frac{\mathrm{d}f_L(l_\mathrm{p}\eta)}{\mathrm{d}l_\mathrm{p}} = f_L(l_\mathrm{p}\eta)\left(a \ln{\frac{l_\mathrm{p}\eta}{l_\mathrm{min}}} + a + \frac{b}{l_\mathrm{p}\eta}\right)\eta.
\end{align}
Therefore, finding the PDF of $L_\mathrm{p}$ requires calculating three different integrals over $\eta$:
\begin{align}
    f_{L_\mathrm{p}}(l_\mathrm{p}) = &-(1+b)\int_1^\infty \sqrt{1 - \frac{1}{\eta^2}}\ f_L(l_\mathrm{p} \eta)\ \mathrm{d}\eta\nonumber\\
    &-l_\mathrm{p}a\left(1+\ln{\frac{l_\mathrm{p}}{l_\mathrm{min}}}\right)\int_1^\infty \sqrt{\eta^2-1}f_L(l_\mathrm{p}\eta)\ \mathrm{d}\eta\nonumber\\
    &-l_\mathrm{p}a\int_1^\infty \sqrt{\eta^2-1}f_L(l_\mathrm{p}\eta)\ln{\eta}\ \mathrm{d}\eta\ \ \ \ \text{for } l_\mathrm{p} > l_\mathrm{min}.
\end{align}
The PDF of $L_\mathrm{p}\ \vert\ L_\mathrm{p} \geq l_\mathrm{p,GRG}$ follows through Eq.~\ref{eq:projectedLengthPDFGRG}.

\section{Likelihood trick}
\label{app:likelihood}
Thanks to its Poissonian form, there exists a particularly numerically efficient way of computing the likelihood presented in Sect.~\ref{sec:theoryInference} as a function of $n_\mathrm{GRG}$, for fixed values of the other parameters.
Defining
\begin{align}
    A(\vec{\theta}) \coloneqq \sum_{i=1}^{N_\mathrm{b}} N_i \ln{\lambda_i(\vec{\theta})}\ \ \ \text{and}\ \ \ B(\vec{\theta}) \coloneqq \sum_{i=1}^{N_\mathrm{b}}\lambda_i(\vec{\theta}),
\end{align}
one interested in the log-likelihood up to a constant only needs to compute
\begin{align}
    \ell(\vec{\theta}) \coloneqq \ln{\mathcal{L}(\{N_i\}\ \vert\ \vec{\theta})} + \sum_{i=1}^{N_\mathrm{b}}\ln{(N_i!)} = A(\vec{\theta}) - B(\vec{\theta}).
\end{align}
The quantity $B(\vec{\theta})$ has a simple interpretation: it is the total number of giants expected to be observed under $\vec{\theta}$ within the entire projected length--redshift parameter space considered.

How does $\ell$ change upon changing $n_\mathrm{GRG}$?
When $n_\mathrm{GRG} \mapsto a \cdot n_\mathrm{GRG}$, $\lambda_i \mapsto a \cdot \lambda_i$, so that
\begin{align}
    \ell(n_\mathrm{GRG}) &\mapsto \sum_{i=1}^{N_\mathrm{b}} N_i \ln{(a \cdot \lambda_i)} - a \cdot \lambda_i\nonumber\\
    &= A(n_\mathrm{GRG}) - a \cdot B(n_\mathrm{GRG}) + \ln{a} \cdot \sum_{i=1}^{N_\mathrm{bins}} N_i.
\end{align}
(In the notation $\ell(n_\mathrm{GRG})$, $A(n_\mathrm{GRG})$, and $B(n_\mathrm{GRG})$, we suppress the dependence on the other five parameters.)
We conclude that, when $n_\mathrm{GRG}$ increases by a factor $a$, the $A$-term in $\ell$ remains the same, the $B$-term in $\ell$ becomes a factor $a$ bigger, and an extra factor emerges: namely, the product of $\ln{a}$ and the total number of giants in the data set.

The significance of this result is that, once $A$ and $B$ are known at some reference number density $n_\mathrm{GRG,ref}$, we can rapidly evaluate $\ell$ for any other number density.
In this work, we implement this `likelihood trick' by evaluating $\ell$ for two different values of $n_\mathrm{GRG}$ (and for many different values of the other parameters).
We then solve for $A(n_\mathrm{GRG,ref})$ and $B(n_\mathrm{GRG,ref})$, and use
\begin{align}
    \ell(n_\mathrm{GRG}) &= A(n_\mathrm{GRG,ref}) - \frac{n_\mathrm{GRG}}{n_\mathrm{GRG,ref}} \cdot B(n_\mathrm{GRG,ref})\nonumber\\
    &+ \ln{\frac{n_\mathrm{GRG}}{n_\mathrm{GRG,ref}}} \cdot \sum_{i=1}^{N_\mathrm{b}}N_i.
\end{align}

\section{PyBDSF parameters}
\label{app:pybdsf}
As described in Sect.~\ref{sec:detection}, the GRG detection pipeline uses PyBDSF for the initial radio blob detection.
For reproducibility, we provide the specific parameters used, which we adopted from \citet{Shimwell2022}:
\begin{verbatim}
bdsf.process_image(<filename>,thresh_isl=4.0,
thresh_pix=5.0, rms_box=(150,15), rms_map=True, 
mean_map='zero', ini_method='intensity', 
adaptive_rms_box=True, adaptive_thresh=150, 
rms_box_bright=(60,15), group_by_isl=False, 
group_tol=10.0, output_opts=True, atrous_do=True, 
atrous_jmax=4, flagging_opts=True, 
flag_maxsize_fwhm=0.5, advanced_opts=True, 
blank_limit=None, frequency=143.65e6)
\end{verbatim}
\FloatBarrier

\section{Adaptations of the radio ridgeline based host galaxy identification}
\label{app:ridge}

\begin{figure*}
\includegraphics[width=\textwidth]{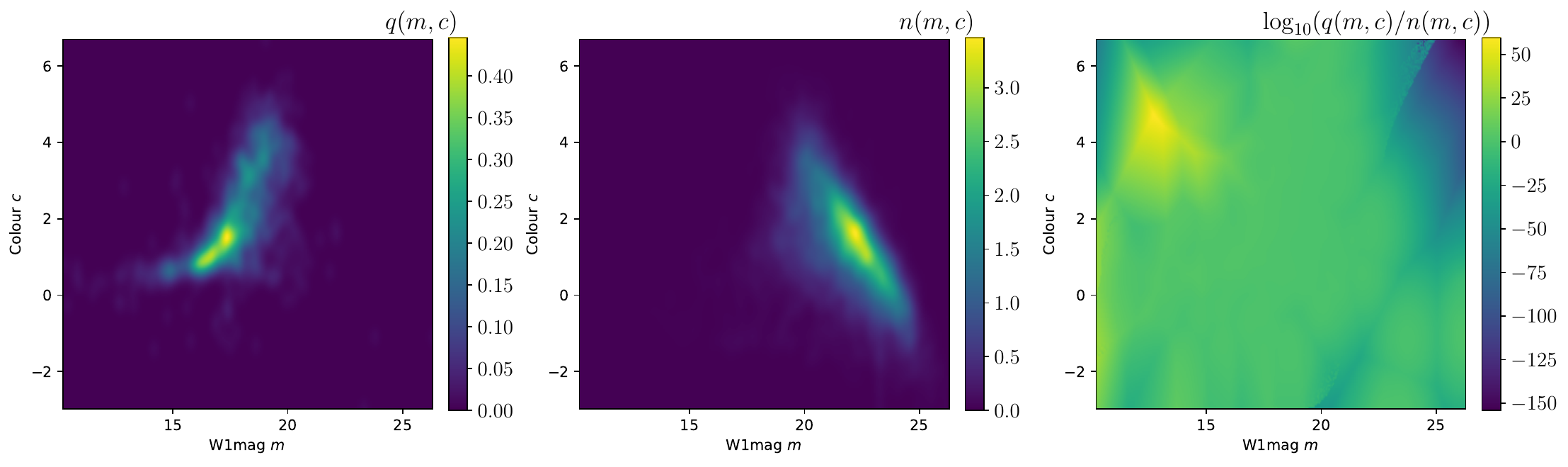}
\caption{Unregularised KDE estimates for $q$ in the left panel, $n$ in the second panel, and $q/n$ with logarithmic colour bar in the third panel.
The KDE bandwidth of $0.2$ stems from \citet{Barkus2022}.}
\label{fig:unregularised}
\end{figure*}

\begin{figure*}
\includegraphics[width=\textwidth]{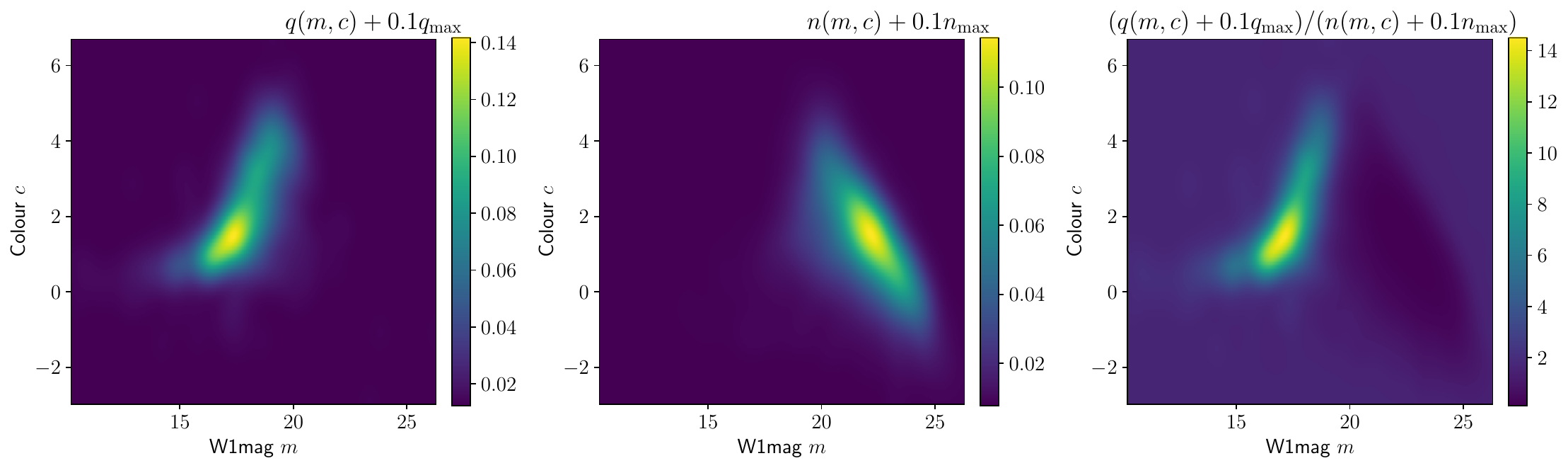}
\caption{Regularised KDE estimates for $q$ in the left panel, $n$ in the second panel, and $q/n$ with logarithmic colour bar in the third panel.
The KDE bandwidth of $0.4$ stems from 10-fold cross-validation.}
\label{fig:regularised}
\end{figure*}

Here we elaborate on two small adaptations of the radio--optical crossmatch method introduced by \citet{Barkus2022}.
First, we explicitly regularised $q(m,c)$ and $n(m,c)$.
Figure \ref{fig:unregularised} shows that the unregularised forms of $q$ and $n$  can take on extreme values in the $LR$ (eq. \ref{eq:LR}) in sparsely sampled regions of the $(m,c)$--parameter space.
The 2D KDE that models $q(m,c)$ was fitted on the $m$ and $c$ values of all $905$ sources with an angular length  $\phi > 1'$ from $40$ randomly picked LoTSS DR2 pointings. The 2D KDE that models $n(m,c)$ was fitted on the $m$ and $c$ values of $10,000$ sources that were randomly sampled from the entire combined infrared--optical catalogue.
By simply adding a small constant factor to $q(m,c)$ and $n(m,c)$ we get more robust $LR$ values, see Fig. \ref{fig:regularised}.
We added a constant factor $0.1\cdot q_\mathrm{max}$ and $0.1\cdot n_\mathrm{max}$ for $q$ and $n$ respectively, where $q_\mathrm{max}$ is the maximum of the KDE for $q$ and $n_\mathrm{max}$ is the maximum of the KDE for $n$.
We set the KDE bandwidths to $0.4$ following a 10-fold cross-validation.

\begin{figure*}
\includegraphics[width=\textwidth]{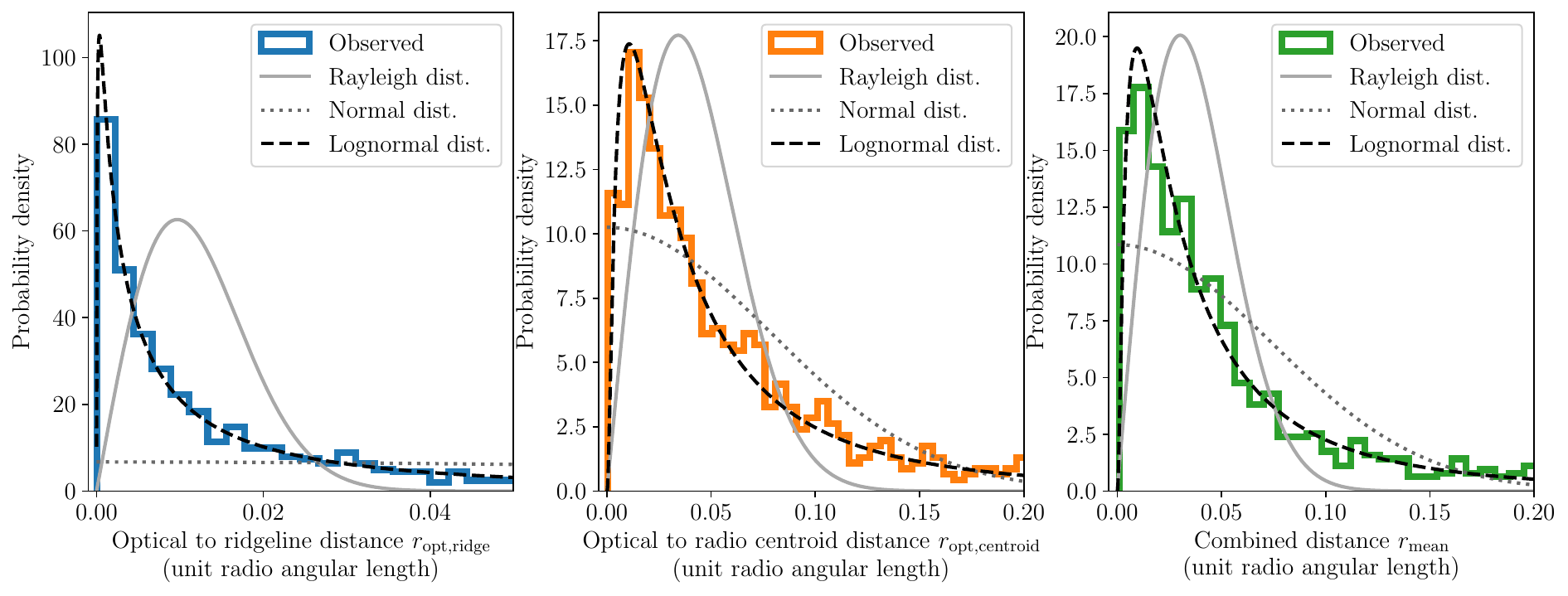}
\caption{Each panel shows the histogram of a different distance measure between $905$ radio galaxies with $\phi > 1'$ and their optical host. The grey, dark grey, and black lines show empirical fits to these histograms for Rayleigh, normal, and lognormal distributions respectively.
The tails of the histograms are long; for visualisation purposes we only plot the $r$-axis up to $0.05$, $0.20$, and $0.20$.}
\label{fig:f(r)}
\end{figure*}

Second, we changed the form of $f(r)$.
Theoretically, we might expect both the distance between the `true' optical counterpart and the radio ridgeline $r_\mathrm{opt,ridge}$ and the distance between the `true' optical counterpart and the radio centroid $r_\mathrm{opt,centroid}$ to be Rayleigh distributed.\footnote{
In two dimensions, the Euclidean distance between the origin and a point whose Cartesian coordinates are independent, zero-mean, and equal-variance normal random variables, is Rayleigh distributed.
This motivates modelling the angular distance between the optical counterpart and the radio centroid with a Rayleigh distribution.
The appropriate value of the distribution's parameter likely depends (positively) on the angular length of the RG considered; as such, one would not expect a single Rayleigh distribution to work for all RGs.
}
However, as Fig.~\ref{fig:f(r)} demonstrates, the lognormal distribution clearly provides the best empirical fit to the distances.
The figure shows a histogram of the distance measures for RGs to their optical counterpart as manually identified through RGZ.
Specifically, we plot the distances for the same $905$ RGs, with an angular length $\phi > 1'$, from $40$ randomly selected pointings as above.
Thus we update $f(r)$ to be:
\begin{align}
    f(r_\mathrm{mean}) = \frac{1}{r_\mathrm{mean} \sigma \sqrt{2 \pi}} e^{(-\frac{(\ln{r_\mathrm{mean}}-\mu)^2}{2 \sigma^2})},
\end{align}
where we empirically determined $\sigma$ and $\mu$ using our sample of $905$ RGs, 
\begin{align}
    \mu=\frac{\sum_i{\ln{r_{\mathrm{mean},i}}}}{n}=-3.37
\end{align} and 
\begin{align}
    \sigma^2=\frac{\sum_i{(\ln{r_{\mathrm{mean},i}}-\mu)^2}}{n}=1.28,
\end{align} with $n=905$ the size of our sample.
\FloatBarrier

\section{Sky coverages}
\label{app:sphericalQuadrangles}
As an extension of Sect.~\ref{sec:methods_parameters}, this appendix details the sky coverages of our analyses.
In particular, Table~\ref{tab:sphericalQuadrangles} provides a decomposition -- in terms of disjoint spherical quadrangles -- of the sky coverage common between the ML pipeline, RGZ, and the combined manual search of \citet{Dabhade2020March} and \citet{Oei2023Distribution}.
For simplicity, and as an acknowledgement of the wiggle room inherent to defining this joint sky coverage, we chose integer coordinates.
Together, these spherical quadrangles cover $\Omega = 5327.9\ \mathrm{deg}^2 = 1.62\ \mathrm{sr}$ (25.8\%) of the Northern Sky.
We refer to this coverage simply as the `LoTSS DR2 coverage'.

The RGZ--ML--\citet{Oei2023Distribution} overlap region amounts to the LoTSS DR2 coverage with the LoTSS DR1 spherical quadrangle removed.
The minimum and maximum right ascensions of this quadrangle are $\alpha_\mathrm{min} = 160\degree$ and $\alpha_\mathrm{max} = 230\degree$, while its minimum and maximum declinations are $\delta_\mathrm{min} = 45\degree$ and $\delta_\mathrm{max} = 56\degree$.
This smaller overlap region covers $4838.9\ \mathrm{deg}^2$ (23.5\%) of the Northern Sky.
It is the sky coverage relevant to estimating the identification probability functions of Sect.~\ref{sec:empirical} and Fig.~\ref{fig:probabilityObservingID}: $p_\mathrm{obs,ID,1}(l_\mathrm{p},z)$, $p_\mathrm{obs,ID,2}(l_\mathrm{p},z)$, and $p_\mathrm{obs,ID}(l_\mathrm{p},z)$.

\begin{table}[]
\caption{
Sky coordinates and solid angles of disjoint spherical quadrangles whose union forms the LoTSS DR2 sky coverage -- over which we have performed our inference.
For each spherical quadrangle, we provide the minimum and maximum right ascension, $\alpha_\mathrm{min}$ and $\alpha_\mathrm{max}$, the minimum and maximum declination, $\delta_\mathrm{min}$ and $\delta_\mathrm{max}$, and its solid angle, $\Omega$.
We list the largest quadrangles first.
The second and third quadrangle touch along the $360\degree$--$0\degree$ right ascension coordinate boundary, and could be viewed as a single whole.
}
\centering
\begin{tabular}{l l l l l}
\hline\hline
$\alpha_\mathrm{min}\ (\degree)$ & $\alpha_\mathrm{max}\ (\degree)$ & $\delta_\mathrm{min}\ (\degree)$ & $\delta_\mathrm{max}\ (\degree)$ & $\Omega\ (\mathrm{deg}^2)$\\\hline
$120$ & $253$ & $28$ & $69$ & $3536.7$\\
$0$ & $35$ & $16$ & $35$ & $597.5$\\
$338$ & $360$ & $16$ & $35$ & $375.6$\\
$253$ & $269$ & $28$ & $47$ & $240.1$\\
$109$ & $120$ & $25$ & $41$ & $147.1$\\
$269$ & $277$ & $31$ & $47$ & $99.2$\\
$330$ & $338$ & $17$ & $30$ & $95.2$\\
$191$ & $210$ & $23$ & $28$ & $85.7$\\
$35$ & $41$ & $24$ & $32$ & $42.3$\\
$253$ & $260$ & $58$ & $69$ & $34.3$\\
$120$ & $131$ & $25$ & $28$ & $29.5$\\
$277$ & $281$ & $41$ & $47$ & $17.3$\\
$327$ & $330$ & $17$ & $20$ & $8.5$\\
$277$ & $280$ & $32$ & $35$ & $7.5$\\
$117$ & $120$ & $53$ & $57$ & $6.9$\\
$260$ & $264$ & $66$ & $69$ & $4.6$\\
\hline
\end{tabular}
\label{tab:sphericalQuadrangles}
\end{table}

\end{appendix}
\end{document}